\documentclass{aa}  
\usepackage{graphicx}
\usepackage{txfonts}
\usepackage[colorlinks=false]{hyperref}
\usepackage{mathtools}
\usepackage{ragged2e}

\begin{document} 
   \title{Eccentric discs as a gateway to giant planets outward migration}
   \author{C. E. Scardoni
          \inst{1}
          \and
          G. P. Rosotti
          \inst{1}
          \and
          C. J. Clarke
          \inst{2}
          \and
          E. Ragusa
          \inst{1}
          \and
          R. A. Booth
          \inst{3}
          }
   \institute{Dipartimento di Fisica, Università degli Studi di Milano, Via Celoria 16, Milano, 20133, Italy\\
              \email{chiara.scardoni@unimi.it}
         \and
             Institute of Astronomy, University of Cambridge, Madingley Road, Cambridge, CB3 0HA, UK
         \and
             School of Physics and Astronomy, University of Leeds, Leeds, LS2 9JT, UK
             }

   \date{Received XXX; accepted XXX}

  \abstract
   {Recent studies on the planet-dominated regime of Type II migration demonstrated the presence of a correlation between the direction of massive planet migration and the parameter $K$ describing the depth of the gap opened by the planet. Indeed, it was found that high (low) value for the $K$ parameter correspond to outward (inward) migration.}
   {In this paper we aim at understanding the mechanism driving inward and {outward} migration and why it correlates with the gap depth.}
   {We performed a suite of 2D, live-planet, long-term simulations of massive planets migrating in discs with the hydro-code {\sc Fargo3D}. We focus on a range of planet masses (1–13 $m_{\rm J}$) and disc aspect ratios (from 0.03 to 0.1), and analyze the evolution of orbital elements and gap structure. We also study the torque contributions from outer Lindblad resonances to investigate their role in the migration outcome.}
   {{We find that, while all planets initially migrate inwards, those with high enough  {$K$} eventually enter a phase in which the torque reverses sign and migration becomes outwards, until eventually stalling.} This behavior is associated with eccentricity growth in the outer disc and changes in the gap structure. We identify the surface density ratio at the 1:2 and 1:3 outer Lindblad resonances as a key output diagnostic that correlates with the migration direction. In general, this ratio regulates the migration for all the cases where the massive planet remains in  {an} almost circular orbit and the outer gap region exhibits moderate eccentricity. {This characteristic sequence of inward-reversal-outwards-stalling can occur for a variety of $K$ values and therefore further work is required to identify the simulation input parameters that determine the onset of this sequence.}}
   {Our results suggest that outward migration in the planet-dominated regime is primarily governed by the relative importance of the 1:2 and 1:3 resonances and, therefore, the gap profiles plays a crucial role in determining the direction of migration.}


   \maketitle

\section{Introduction}
Once formed in protoplanetary discs, planets interact gravitationally with the surrounding gas by exchanging angular momentum and orbital energy (e.g. \citealt{Lin&Papaloizou1993,Goldreich&Tremaine1979,Goldreich&Tremaine1980} and recent reviews by \citealt{Papaloizou&Terquem2006,Kley&Nelson2012,Baruteau+2014,Papaloizou2021,Paardekooper+2022}). This planet-disc interaction alters the disc structures by forming, for example, spirals and gaps (e.g. \citealt{Papaloizou&Lin1984, Hourigan&Ward1984, Kley&Nelson2012}), and determines the evolution of the planet orbital parameters (e.g. \citealt{Ward1997}). Since it modifies the planet semi-major axis and eccentricity, disc-planet interaction is therefore a key process shaping the architecture of planetary systems (e.g. \citealt{Ida&Lin2008, Mordasini2018}). Understanding the physical mechanisms that govern the direction and rate of this migration is thus essential to explain the observed distribution of exoplanets and to connect planet formation models with observations of young discs (e.g., \citealt{Andrews2018,Nielsen+2019,Long+2018}).

Depending on the planet’s mass and the disc properties, migration is typically divided into two regimes \citep{Artymowicz1993,Ward1998,Kley&Nelson2012}: Type I migration for low-mass, embedded planets, and Type II migration for massive planets that open deep gaps (e.g., \citealt{Ward1997, Ward1986,Lin&Papaloizou1993,Syer&Clarke1995,Ivanov+1999}). According to the classical picture, once the planet opens a gap and enters the Type II migration regime, no disc material can cross the planetary gap (e.g., \citealt{Papaloizou&Lin1984,Lin&Papaloizou1993,Ida&Lin2008,Bitsch+2013}). As a consequence, if the local disc is more massive than the planet ('disc-dominated regime') the planet is locked to the viscous evolution of the disc and behaves as a gas particle; otherwise, when the planet becomes more massive than the local disc ('planet-dominated regime') its inertia is so high that it significantly slows down the migration  {(\citealt{Syer&Clarke1995} and \citealt{Ivanov+1999})}. These two regimes of Type II migration are typically identified via the disc-to-planet mass ratio parameter ($\Sigma(r_{\rm p})$ is the disc density at the planet location and $m_{\rm p}$ is the planet mass)
\begin{equation}
    B=\frac{4\pi \Sigma(r_{\rm p}) r_{\rm p}^2}{m_{\rm p}}.
\end{equation}

However, this classic view of Type II migration has been challenged by recent numerical simulations showing that the disc material can cross the planetary gap \citep{Lubow&Dangelo2006}, and this unlocks the planet from the disc viscous evolution \citep{Duffell+2014,Duermann&Kley2015}. Several studies on the disc-dominated regime of Type II migration have investigated this issue (e.g., \citealt{Kanagawa+2018, Robert+2018,Scardoni+2020,Scardoni+2022, Lega+2021}). Some suggest that planet migration is independent of the disc viscous evolution \citep{Duermann&Kley2015,Kanagawa+2016,Kanagawa+2018}, while others find that the migration rate remains proportional to the disc viscous evolution \citep{Robert+2018}. It has also been demonstrated that the classical Type II rate is restored if the system has enough time to adjust to the presence of a migrating planet \citep{Scardoni+2020}.

A critical problem related to migration is the study of the planetary gap's properties; as the strongest torque is exerted close to the planet, the details of the gap shape strongly influence the migration outcome {\citep{Goldreich&Tremaine1979, Goldreich&Tremaine1980,Goldreich&Sari2003}}. Although the detailed description of the gap is not straightforward, some works focused on this issue. For example, \citet{Crida+2006} provided an analytical formula linking the gap properties to the planet and disc parameters; \citet{Fung+2014} and \citet{Fung&Chiang2016} studied the gap density and showed that planetary gaps produced in 3D simulations are consistent with those in 2D; \citet{Kanagawa+2015}, instead, studied the gap shape and provided an empirical estimate of the gap width.

In this paper, we focus on the planet-dominated regime of Type II migration (low $B$), which gained popularity recently, with the discovery that in this regime outward migration can be easily achieved in simulations \citep{Dempsey+2020,Dempsey+2021,Scardoni+2022}. In particular, these studies showed that the depth and width of gap opened by the planet play a key role in determining the migration behavior, finding a correlation between the direction of migration and the gap-depth parameter \citep{Kanagawa+2015}
\begin{equation}
    K=\frac{q^2}{\alpha h^5},
    \label{eq:K}
\end{equation}
with planets migrating outward (inward) for high (low) values of $K$. However, so far, such correlation has only been found empirically \citep{Dempsey+2020,Dempsey+2021} and the consequences for the distribution of exoplanets explored \citep{Scardoni+2022}, while we still miss a physical explanation for this correlation.

Building on our previous work \citep{Scardoni+2022}, here our long-term integrations (ranging from 300k to 600k orbits) reveal that in some cases a given planet exhibits a sequence of evolutionary phases: inward migration, torque reversal, outward migration and stalling. 

In \sectionautorefname~\ref{sec:Migration properties of Jupiter mass-like planets}, we examine this behavior in detail in the case of a 1 $m_{\rm J}$ planet with aspect ratio $h=H/r=0.05$, finding that torque reversal is associated with a change in gap structure and an episode of eccentricity growth in the disc exterior to the planet. {We discuss the possible origin of this behavior in terms of planet-disc interaction at the 1:3 resonance; indeed, when the 1:3 resonance becomes dominant over the 1:2 resonance, it can significantly alter the gap structure by increasing the disc eccentricity in the outer gap areas - this can therefore impact the direction of migration.} 
{Indeed, the interaction between the disc and the planet at the outer Lindblad resonances can cause eccentricity growth in the gap area  {(e.g., \citealt{Artymowicz1993, Kley&Dirksen2006, Lubow1991, Papaloizou2021, Tanaka+2002, Tanaka+2022})}. In particular, the 1:2 outer Lindblad resonance (located at $r\sim 1.59\ r_{\rm p}$) tends to damp the eccentricity, while the 1:3 resonance (located at $r\sim 2.08\ r_{\rm p}$) tends to increase the eccentricity.

We show that the change in gap shape at torque reversal can be simply parameterized in terms of the surface density ratio at the 1:2 and 1:3 resonances. We furthermore show across all simulations that the value of this ratio, derived as a simulation {\it output}, determines whether planets migrate inwards, outwards or stall.

In \sectionautorefname~\ref{subsec:Normaliszed torques} we instead attempt to map migration outcomes on to simulation {\it input} parameters, as in \citet{Scardoni+2022}. We, however, find that this sequence of inward migration, torque reversal and stalling occurs over a wide range of values of $H/r$ ($0.04$ to $0.055$ for Jupiter mass planets) corresponding, for the flaring law considered, to around a factor 5 in orbital radius. We expect planets that are located within this range of $H/r$ at the point of entering the planet-dominated regime ($B<1$) to change their orbital radii by no more than a couple of 10s of percent in the  $B<1$ regime. Outside this range we find that at large radius ($H/r>0.06$) migration is inwards over the duration of the experiment, while at smaller radius {($H/r<0.036$)} planets migrate outwards throughout the simulation.

We also examine the migration behavior of more massive planets (3 $m_{\rm J}$ and 13 $m_{\rm J}$; see \sectionautorefname~\ref{sec:Migration of higher mass planets}) and define the limits of validity for the proposed migration mechanism. In particular, we find that migration is regulated by the 1:2 and 1:3 density ratio only when the planet eccentricity remains moderate and disc eccentricity growth is confined near the outer gap edge. In cases of high planet eccentricity or eccentricity excitation over a broad region of the disc, different mechanisms appear to dominate the migration behavior.

This paper is organized as follows: in \sectionautorefname~\ref{sec:Simulations} we introduce our set of live-planet, long-term simulations, with orbital parameters described in \sectionautorefname~\ref{sec:Orbital parameters}. In \sectionautorefname~\ref{sec:Migration properties of Jupiter mass-like planets} and \sectionautorefname~\ref{sec:Migration of higher mass planets}, we present the migration results for Jupiter-mass and higher-mass planets, respectively. Finally, in \sectionautorefname~\ref{sec:Discussion}, we discuss implications of our findings and explore alternative criteria to predict the direction of planet migration.

\section{Simulations}
\label{sec:Simulations}
\subsection{Simulation parameters}
In addition to the set of 12 simulations already presented in \citet{Scardoni+2022}, for this work we used 16 additional 2D hydrodynamical simulations performed with the grid code {\sc Fargo 3D} \citep{BenitezLlambay&Masset2016}. As in the previous work, we adopted a cylindrical reference frame $(r,\varphi)$ centered on the star and considered the dimensionless units $G=M_*=r_0=1$, where $G$ is the gravitational constant, $M_*$ is the star mass and $r_0$ is the planet's initial location. In the adopted units, $\Omega_{\rm k}^{-1}(r_0)=1$ and thus one orbit at the initial planet location $r_0$ is completed in $t=2\pi$. All new simulations are run for at least $10 t_{\nu,0}$ to ensure that they reach condition $|\dot{M}(r)-\dot{M}(r_{\rm in})|/\dot{M}(r_{\rm in}) \leq10\%$ at least until $r=2.5\ a_{\rm p}$ (where $a_{\rm p}$ is the planet's semi-major axis, $t_{\nu,0}$ is the viscous timescale at the initial planet radius and $r_{\rm in}$ is the inner disc radius).

The new simulations' set up is the same as that used in \citet{Scardoni+2022}.\footnote{In analogy with \citet{Ragusa+2018,Rosotti+2017}} We thus consider $N_{r}=430$ logarithmically spaced cells in the radial direction ranging from $r_{\rm in}=0.2$ to $r_{\rm out}=15$; in the azimuthal direction (from $0$ to $2\pi$) we take $N_{\varphi}=580$ linearly spaced cells. We assume that the disc is locally isothermal, with a disc aspect ratio defined as
\begin{equation}
    h=H/r=h_0 r^{f},
    \label{eq:h}
\end{equation}
where $f=0.215$ is the flaring index and $h_0$ is the aspect ratio at the initial planet location.
The viscosity is parameterized using the $\alpha$ prescription of \citet{Shakura&Sunyaev1973}, $\nu=\alpha c_{\rm s} H$, with $\alpha$ varying with the radius as follows
\begin{equation}
    \alpha = \alpha_0 r^{a},
\end{equation}
with $\alpha_0=0.001$ and $a=-0.63$ for all the simulations.

In the previous work, we considered 3 different planet masses ($m_{\rm p}=1\ m_{\rm Jup}$, $m_{\rm p}=3\ m_{\rm Jup}$, $m_{\rm p}=13\ m_{\rm Jup}$), in this work we perform additional simulations with $m_{\rm p}=1\ m_{\rm Jup}$ and $m_{\rm p}=3\ m_{\rm Jup}$ (where $m_{\rm Jup}$ is the mass of Jupiter). 

As in \citet{Scardoni+2022}, we choose the initial disc surface density at the planet location $\Sigma_0$ in a way that we have $B_0=0.15$ for the `massive' disc simulations, and $B_0=0.046$ for the `light' disc simulations (see next section for the initial density profiles).\footnote{For reference, if we consider a Jupiter mass planet at the radius of Jupiter, $B_0=0.046$ and $B_0=0.15$ correspond in physical units to a local disc surface density of $1.3\ \rm g/cm^2$ and  $4.2\ \rm g/cm^2$, respectively.} Note that we only consider values $B_0<1$ because in this work we are interested in the planet-dominated regime of Type II migration.
Apart from $m_{\rm p}$, $h_0$ and $\Sigma_0$, all the other parameters are kept fixed among all the simulations; all the simulations used in this work are summarized in \tableautorefname~\ref{tab:simparameters}.
\begin{table}
    \centering
    \caption{Simulation parameters.}
	\label{tab:simparameters}
    \begin{tabular}{lccccccccc}
        \hline
        {Name} & $B_0$ & $m_{\rm p}[m_{\rm J}]$ & $h_0$ & $K_0$\\
        \hline
        L-m1-h03 & $0.046$ & $1$ & $0.03$  & $4.12 \cdot 10^4$ \\
        L-m1-h036 & $0.046$ & $1$ & $0.036$  & $1.65 \cdot 10^4$ \\
        L-m1-h04 & $0.046$ & $1$ & $0.04$  & $9.77 \cdot 10^3$ \\
        L-m1-h045 & $0.046$ & $1$ & $0.045$  & $5.42 \cdot 10^3$ \\
        L-m1-h05 & $0.046$ & $1$ & $0.05$  & $3.20 \cdot 10^3$ \\
        L-m1-h055 & $0.046$ & $1$ & $0.055$  & $1.99 \cdot 10^3$ \\
        L-m1-h06 & $0.046$ & $1$ & $0.06$  & $1.29 \cdot 10^3$ \\
        L-m1-h065 & $0.046$ & $1$ & $0.065$  & $8.62 \cdot 10^2$ \\
        L-m3-h036 & $0.046$ & $3$ & $0.036$  & $1.49 \cdot 10^5$ \\
        L-m3-h05 & $0.046$ & $3$ & $0.05$  & $2.88 \cdot 10^5$ \\
        L-m3-h06 & $0.046$ & $3$ & $0.06$  & $1.16 \cdot 10^5$ \\
        L-m13-h036 & $0.046$ & $13$ & $0.036$  & $2.9 \cdot 10^6$ \\
        L-m13-h06 & $0.046$ & $13$ & $0.06$  & $2.17 \cdot 10^5$ \\
        L-m13-h1 & $0.046$ & $13$ & $0.1$  & $1.69 \cdot 10^4$ \\
        M-m1-h03 & $0.15$ & $1$ & $0.03$  & $4.12 \cdot 10^4$ \\
        M-m1-h036 & $0.15$ & $1$ & $0.036$  & $1.65 \cdot 10^4$ \\
        M-m1-h04 & $0.15$ & $1$ & $0.04$  & $9.77 \cdot 10^3$ \\
        M-m1-h045 & $0.15$ & $1$ & $0.045$  & $5.42 \cdot 10^3$ \\
        M-m1-h05 & $0.15$ & $1$ & $0.05$  & $3.20 \cdot 10^3$ \\
        M-m1-h055 & $0.15$ & $1$ & $0.055$  & $1.99 \cdot 10^3$ \\
        M-m1-h06 & $0.15$ & $1$ & $0.06$  & $1.29 \cdot 10^3$ \\
        M-m1-h065 & $0.15$ & $1$ & $0.065$  & $8.62 \cdot 10^2$ \\
        M-m3-h036 & $0.15$ & $3$ & $0.036$  & $1.49 \cdot 10^5$ \\
        M-m3-h05 & $0.15$ & $3$ & $0.05$  & $2.88 \cdot 10^5$ \\
        M-m3-h06 & $0.15$ & $3$ & $0.06$  & $1.16 \cdot 10^5$ \\
        M-m13-h036 & $0.15$ & $13$ & $0.036$  & $2.9 \cdot 10^6$ \\
        M-m13-h06 & $0.15$ & $13$ & $0.06$  & $2.17 \cdot 10^5$ \\
        M-m13-h1 & $0.15$ & $13$ & $0.1$  & $1.69 \cdot 10^4$ \\
        \hline
    \end{tabular}
    \tablefoot{
    The simulations are named according to the following rules: `L' or `M' to indicate a light or massive disc, respectively; `m' followed by a number to indicate the planet mass (measured in Jupiter masses); `h' followed by a number to indicate the aspect ratio. $B_0$ and $K_0=q^2/(\alpha h^5)$ are the disc-to-planet mass ratio and the gap depth parameter both evaluated at the initial planet location. The table contains all the simulations used in this work (both the new ones and those already presented in \citet{Scardoni+2022}).
    }
\end{table}

\subsection{Initial and boundary conditions}
\label{subsec:Initial&BCs}
The initial density profile is defined as
\begin{equation}
    \Sigma(r) = \Sigma_0 r^{-0.3} \cdot {\rm e}^{\left(-{r}/{5}\right)^{1.7}},
\end{equation}
The disc is initially smooth, and the planet mass is gradually inserted in the simulation at $R_0=1$ by progressively increasing its mass during the first {50 orbits}; the planet is then released and allowed to evolve under the planet-disc torques at 50 orbits, when it has reached its full mass.

To minimize any effect from the outer boundary, we set both the velocity and the gas density to zero at $r_{\rm out}$, thus no material enters the simulation ('closed boundary condition'); however, since the simulations are run for a $t\ll t_{\nu}(r_{\rm out})$, the effect of the outer boundary condition is negligible. At the inner boundary, instead, the material velocity is chosen to match the viscous velocity as in \citet{Scardoni+2020, Scardoni+2022} and \citet{Dempsey+2020,Dempsey+2021} ('viscous boundary condition') to ensure that the torque supplied at the inner boundary is equal to the rate of angular momentum advected by accretion (i.e. the total disc angular momentum is not modified). To prevent the propagation to the inner disc of spurious waves generated by boundary effects, we employ the wave damping method by \citet{Val-Borro+2006} for the radial velocity.\footnote{Over the damping timescale ($\tau=\Omega_{\rm k}^{-1}/30$), we damp $v_r$ to the azimuthal average of the initial (viscous) velocity $\langle v_{r,0}\rangle$ in the region from $r_{\rm in}$ to $r=0.3$.}

\section{Orbital parameters}
\label{sec:Orbital parameters}
\figureautorefname~\ref{Fig:Ecc_SemiAxis_Evolution} shows the orbital parameters as a function of the evolutionary time (defined as the ratio between the physical time and the viscous timescale $t/t_{\nu,0}$) for our entire set of simulations up to $10\ t_{\nu,0}$.\footnote{We limit our analysis to $10\ t_{\nu,0}$ to avoid spurious effects from the outer boundary, where the viscous timescale is $\sim 19 t_{\nu,0}$.} The left panels show the light disc simulations ($B=0.046$), while the right panels show the massive disc simulations ($B=0.15$). {We stress that these simulations are evolved for unusually long timescales (from 300k to 600k orbits), enabling us to capture the long-term dynamical behavior of the system, as described below; indeed, the value of the viscous timescale at the initial planet location $t_{\nu,0}$ ranges from $\sim 1.16\cdot 10^5$ orbits (for $h=0.03$) to $\sim 250 \cdot 10^3$ (for $h=0.065$).}

By analyzing the evolution of the semi-major axis, shown in the upper panels of \figureautorefname~\ref{Fig:Ecc_SemiAxis_Evolution}, the simulations reveal a variety of behaviors for planet migration, with planets migrating inward, outward, stalling, or changing the direction of migration during their evolution. The semi-major axis evolution appears slower in the light disc case (left panel) with respect to the massive disc case (right panel), with the evolution time scaling inversely with the disc mass; this behavior is consistent with what expected for planet-dominated type II migration, where the migration rate is slower for lower values of the disc-to-planet mass ratio $B$, since the torques per unit disc mass become independent of disc mass. Overall, outward migration is favored by higher mass planets and lower aspect ratio values.

The lower panels of \figureautorefname~\ref{Fig:Ecc_SemiAxis_Evolution} show the evolution of the planet eccentricity for all the simulations. From these plots, we can notice that 1 $m_{\rm Jup}$ planets never grow eccentricity higher than 0.05; 3 $m_{\rm Jup}$ planets have eccentricities in the range $\sim 0.05-0.13$; 13 $m_{\rm Jup}$ planets, instead, grow eccentricities larger than 0.2 both in the light and in the massive disc case.

\begin{figure*}
    \centering
    \includegraphics[width=1\linewidth]{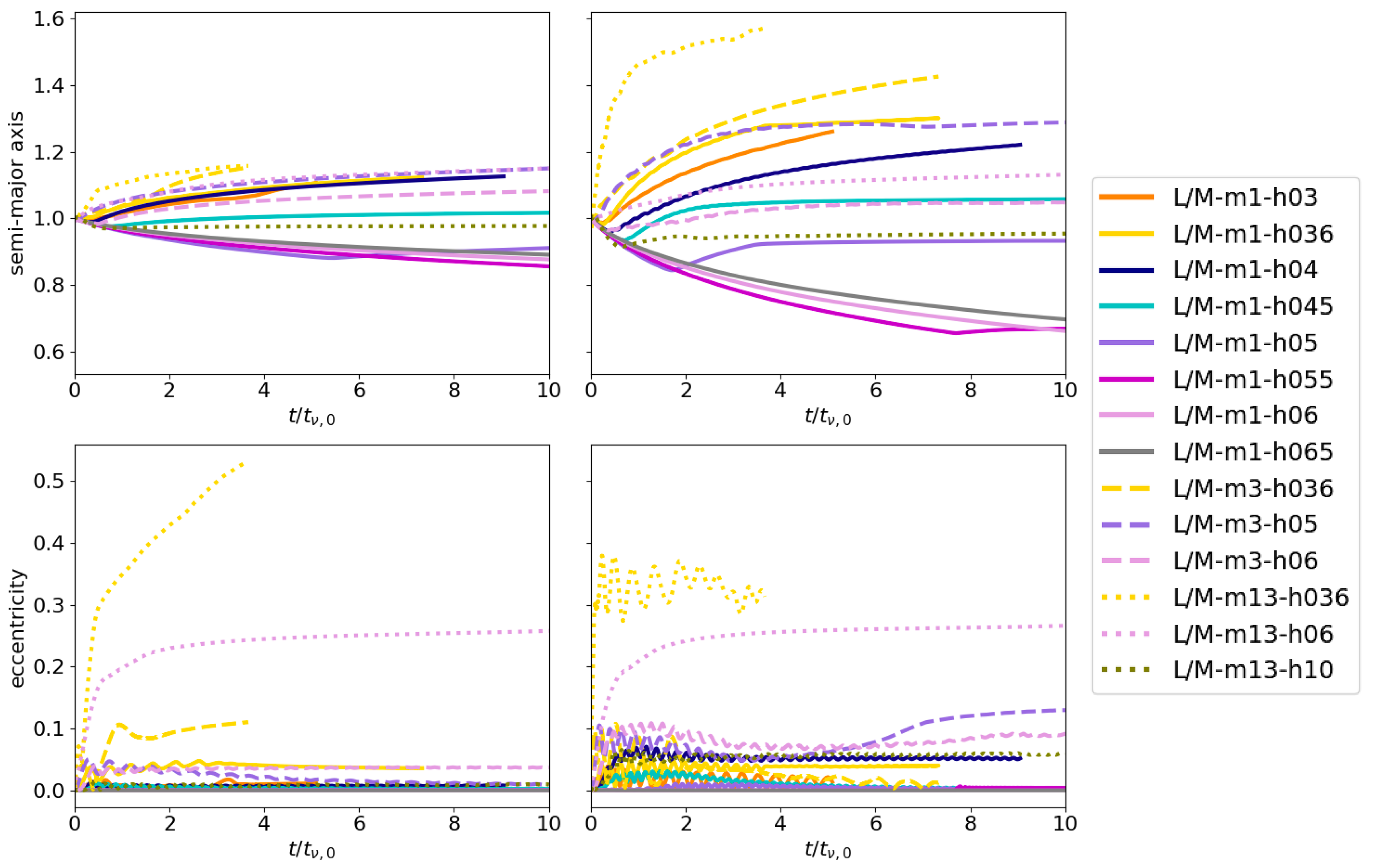}
    \caption{Semi-major axis (upper panels) and eccentricity (lower panels) as a function of the evolutionary time. The left panels show the `light' disc simulations, the right panels show the `massive' disc simulations. As indicated in the legend, the solid/dashed/dotted lines show the orbital parameters for simulations with 1 $m_{\rm Jup}$ / 3 $m_{\rm Jup}$ / 13 $m_{\rm Jup}$ planets. The different colors correspond to different disc aspect ratios.}
    \label{Fig:Ecc_SemiAxis_Evolution}
\end{figure*}

\section{Migration properties of Jupiter mass-like planets}
\label{sec:Migration properties of Jupiter mass-like planets}
Initially, we focus on Jupiter-mass planets because they offer a clean benchmark to study planet migration, thanks to the low eccentricity developed by the planets during the evolution of the system. In this section, we focus on the following questions:
\begin{enumerate}
    \item What is the cause of outward migration?
    \item How do we interpret the migration tracks obtained from the simulations?
    \item Can we predict the direction of migration based on the disc and planet properties?
\end{enumerate}

\subsection{The role of the disc structure and eccentricity in outward planet migration}
\label{subsec:The role of the disc structure and eccentricity in outward planet migration}
In this section, we analyze the disc structure during the evolution of the system, with the goal of understanding how the disc properties influence the direction of migration of the planet. For this purpose, we select as a case study the simulation {M-m1-h05} in which the planet initially migrates inwards, then reverses its direction of migration just before $2\ t_{\nu,0}$ and finally stalls (from $\sim 4\ t_{\nu,0}$).

The disc evolution is illustrated in \figureautorefname~\ref{Fig:dens_ecc_profiles}: the azimuthally averaged disc density (normalized by $\Sigma_0=\Sigma(r=1,t=0)$) and eccentricity are shown in the left and right panels, respectively, as a function of radius (x-axis) and time (y-axis). The white line shows in both plots the planet migration track. From $t=0$ to $t\lesssim 1.7 \ t_{\nu,0}$ the planet migrates inwards. In this period of time, the disc density initially adjusts to the presence of the planet by creating the gap (in the very first part of the evolution) and redistributes the gas through viscosity, while the eccentricity is below 0.1 throughout the disc. Between $1.7 \ t_{\nu,0}\lesssim t \lesssim 3.8 \ t_{\nu,0}$ the planet migrates outwards; meanwhile the outer gap region becomes eccentric and the amount of material passing through the gap increases, causing the inner disc region to become denser.
\begin{figure*}
    \centering
    \includegraphics[width=0.5\linewidth]{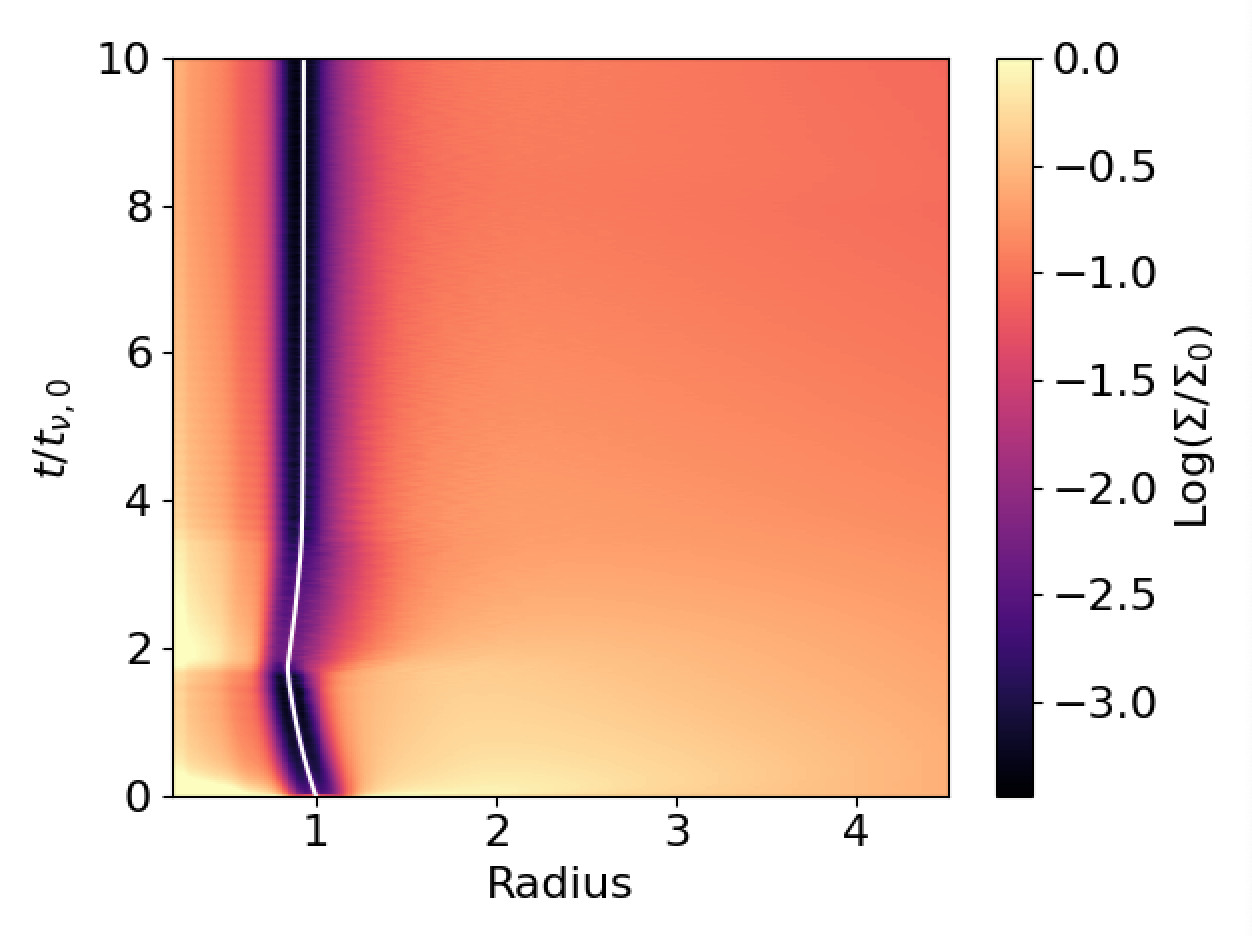}%
    \includegraphics[width=0.5\linewidth]{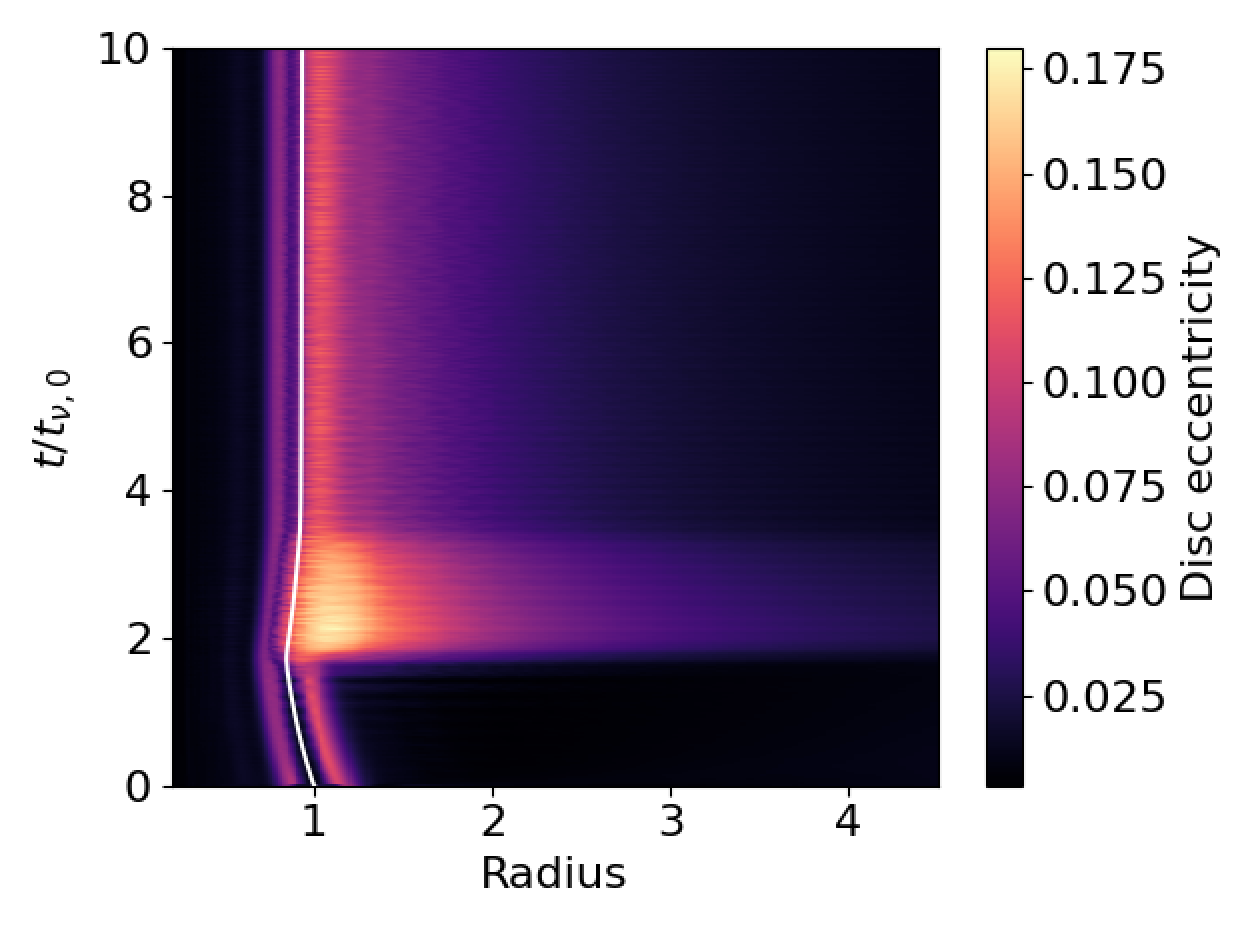} 
    \caption{Simulation M-m1-h05. The colormaps show the azimuthal average of the disc normalized density $\Sigma/\Sigma_0$ (left panel) and of the disc eccentricity (right panel) as a function of the disc radius (x-axis) and of evolutionary time $t/t_{\nu,0}$ (y-axis). The white line in both plots shows the planet migration track.}
    \label{Fig:dens_ecc_profiles}
\end{figure*}

The increase in outer disc eccentricity suggests the presence of a resonant interaction between the disc and the perturber, i.e. the planet. We therefore examine in detail the disc density profiles close to $t\sim 1.7 \ t_{\nu,0}$, where the planet changes its direction of migration, with the goal of understanding whether a resonant interaction is occurring and whether the eccentricity increase is a cause or a consequence of the planet outward migration.

The left panel of \figureautorefname~\ref{Fig:MigrationStudy} illustrates the migration track for the selected simulation M-m1-h05, with an inset {zooming in to} the time interval around the moment when the planet changes its direction of migration, which is marked by the green dot. We also highlight some timesteps during inward migration (yellow and blue dots) and outward migration (pink and gray dots), selected just before and after the change in migration direction. The central-left panel displays the disc density profiles corresponding to each of these timesteps, with line colors matching those of the time markers in the left panel. From the density profiles, we can notice that as the planet switches from inward to outward migration, the density of the inner disc increases (as already observed in \figureautorefname~\ref{Fig:dens_ecc_profiles}), the depth of the gap decreases (coherent with the fact that there is more gas flow through the gap), and the outer gap density decreases (consistent with the outer gap area becoming eccentric). 
The central-right panel shows the planet and disc eccentricity, computed through the angular momentum deficit (AMD), i.e. the difference between the disc element angular momentum and that of an equivalent disc element on circular orbits {(for a detailed computation see, for example, \citealt{Ragusa+2018})}. We notice that both the planet and the disc eccentricity increase just before the change in sign of the direction of migration (the colored lines show the selected snapshots). From this plot, we can notice that the eccentricity starts increasing before the planet changes its direction of migration, suggesting that the change in migration direction might be a consequence -- and not a cause -- of the eccentricity increase. The rapid increase in eccentricity, as previously noted, suggests the presence of a resonant interaction. Eccentricity growth at the outer gap edge can occur when the outer 1:3 Lindblad resonance {becomes} dominant over the outer 1:2 Lindblad resonance {(see, e.g., \citealt{Papaloizou+2001,Goldreich&Sari2003,Tanaka+2002})}. {Furthermore, the simultaneous increase of the AMD of both disc and planet cannot be the result of a secular interaction and therefore points to the importance of resonances at this evolutionary stage.} We thus plot in the right panel of \figureautorefname~\ref{Fig:MigrationStudy} the ratio between the disc density at the 1:2 resonance $\Sigma_{1:2}$ and that at the 1:3 resonance $\Sigma_{1:3}$ (similarly to \citealt{Tanaka+2022}). As can be seen in this plot, while the planet migrates inwards, the density ratio at the two resonances is high $\Sigma_{1:2}/\Sigma_{1:3}\sim 0.75$, and it starts to decrease just before the change in the migration direction. $\Sigma_{1:2}/\Sigma_{1:3}$ remains low for all the outward migration phases, and it increases again when the planet stalls. We verified that also the other simulations show a similar behavior (see \appendixautorefname~\ref{app:Evolution of the 2:1 vs 3:1 resonances}) 
\begin{figure*}
    \centering
    \includegraphics[width=1\linewidth]{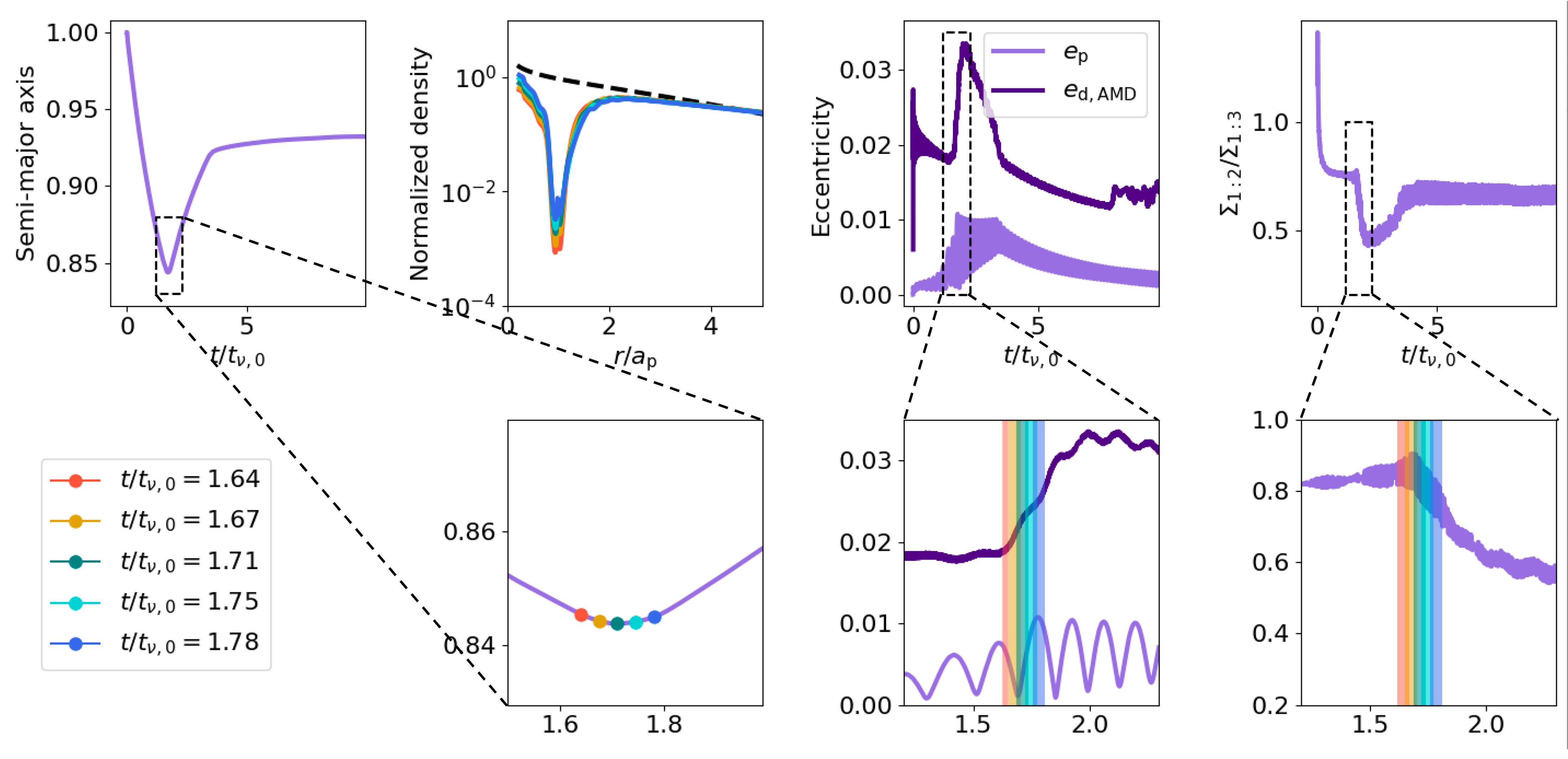}
    \caption{Analysis of the migration behavior in simulation M-m1-h05. The left panel shows the semi-major axis as a function of the evolutionary time, with a zoom in the region where the planet changes its direction of migration, marked via the green dot. The other colored dots mark selected snapshots just before and after the change in migration direction. The central-left panel shows the density profiles at the selected snapshots. The central-right panel shows the disc (dark purple) and planet eccentricity (medium purple) evolution; the colored regions highlight the selected snapshots. The plot in the right panel illustrates the density ratio at the 1:2 and 1:3 resonances.}
    \label{Fig:MigrationStudy}
\end{figure*}

In \figureautorefname~\ref{Fig:TorqueVSres} we show $\Sigma_{1:2}/\Sigma_{1:3}$ on the x-axis, while on the y-axis we show the portion of the normalized torque responsible for the change in the semi-major axis
\begin{equation}
    \Delta T^* = \frac{J_{\rm p}}{2}\cdot \frac{\dot{a_{\rm p}}}{a_{\rm p}},
    \label{eq:DeltaT*}
\end{equation}
where $J_{\rm p}=m_{\rm p}\sqrt{GM_* a_{p}(1-e_{\rm p}^2)}$ is the planet angular momentum.\footnote{Note that the total torque acting on the planet is $\Delta T=\Delta T^* - J_{\rm p} e/(1-e^2)\cdot \dot{e}$} {The different colored lines correspond to the various simulations, as indicated in the legend (with colors matching those in \figureautorefname~\ref{Fig:Ecc_SemiAxis_Evolution}). The initial and final states of each run (averaged over 30,000 orbits) are marked by a colored circle and square, respectively. Each line traces the evolution of the torque as a function of the density at the 1:2 and 1:3 resonances. The inset illustrates the typical trajectory in this plane (shown for our reference simulation M-m1-h05): the runs usually begin with negative torque and high $\Sigma_{1:2}/\Sigma_{1:3}>0.8$, then move toward the upper-left region of the plot—corresponding to low $\Sigma_{1:2}/\Sigma_{1:3}<0.5$ and positive torque—and eventually approach the zero-torque condition at intermediate values $\Sigma_{1:2}/\Sigma_{1:3}\sim 0.6$. Not all simulations complete this loop: some remain at negative torque and high $\Sigma_{1:2}/\Sigma_{1:3}$, while others halt their evolution at positive torque and low $\Sigma_{1:2}/\Sigma_{1:3}$. The level of the evolution depends on the specific disc parameters, in this case the aspect ratio.}
This plot therefore confirms that the 1:3 resonance (and thus the $\Sigma_{1:2}/\Sigma_{1:3}$ ratio) is key in the outward migration process.
\begin{figure*}
    \centering
    \includegraphics[width=1\linewidth]{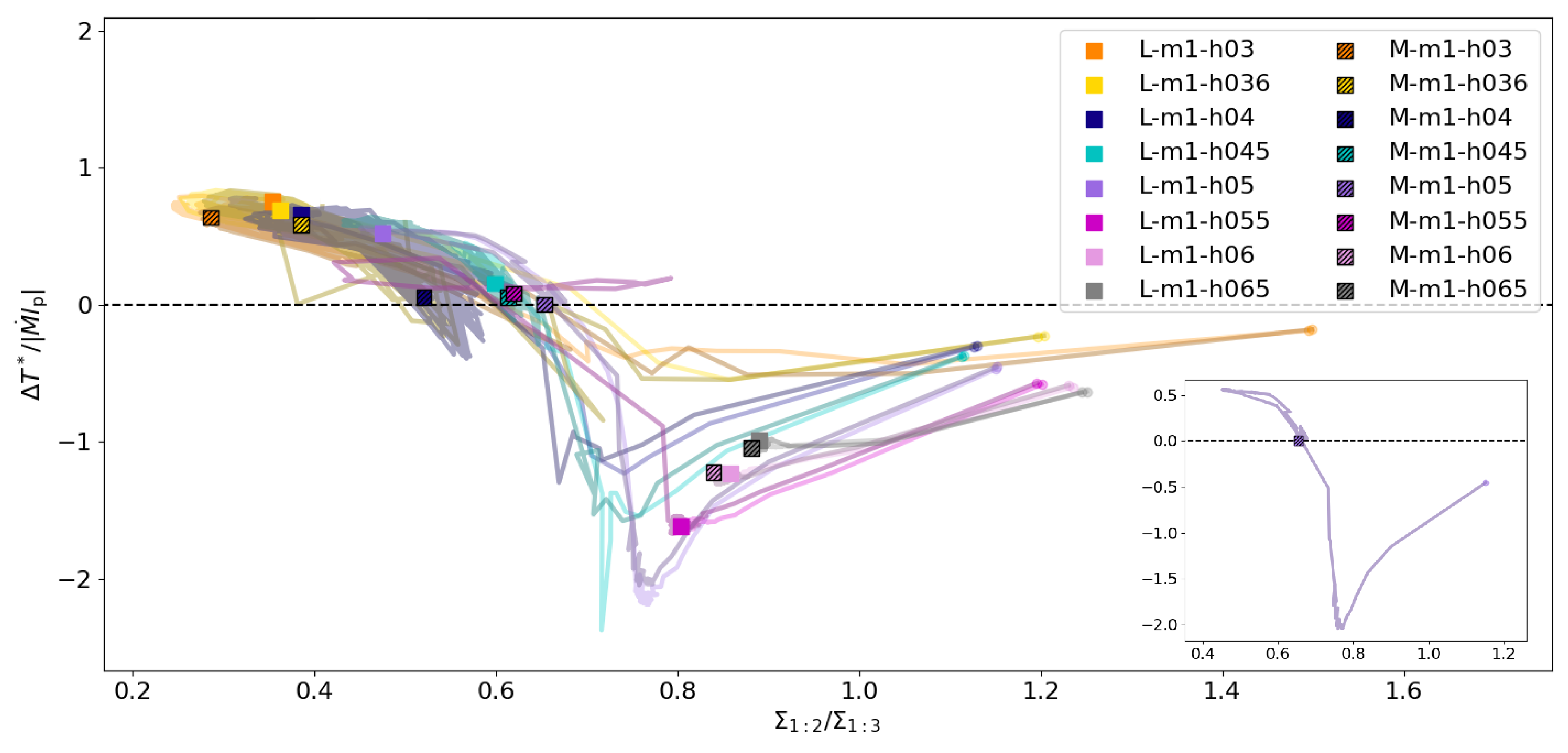}
    \caption{Portion of the normalized torque responsible for the change in the semi-major axis as a function of the density ratio at the 1:2 and 1:3 resonances $\Sigma_{1:2}/\Sigma_{1:3}$. Different colors indicate different disc aspect ratios at 5 au, as indicated in the legend. Markers indicate the state of each simulation at $t=10\ t_{\nu}$. The lines show the evolution of each simulation, whose start is shown by the circles. The inset shows the evolution track for the simulation M-m1-h05. {The high (low) line opacity corresponds to the lower (higher) disc mass regimes.}}
    \label{Fig:TorqueVSres}
\end{figure*} 

As explained by \citet{Papaloizou+2001}, the 1:3 resonance is associated with a rapid increase in eccentricity for the planet and the edge of the outer gap by excitation of an $m=2$ spiral mode. Interestingly, they already suggested that in those cases where the outer gap eccentricity becomes high enough to alter the gap structure, planet outward migration might be triggered. {Indeed, in these cases, due to the different gap shape, the 1:2 resonance (favoring circular disc orbits and inward migration) becomes less important than the 1:3 resonance (favoring eccentric disc orbits and outward migration); this effect can cause the reversal of the direction of migration.} {The physical mechanism and intuition of the relation between the resonances and the direction of migration will be explored further in \sectionautorefname~\ref{subsec:Interpretation of inward and outward migration phases}.} {In particular, they were noticing this effect only for very massive planets, close to the brown dwarf regime; however, lower viscosity and higher integration time in our case might make the effect possible for lower mass planets.} {Another difference from \citet{Papaloizou+2001} is that in the present case there is an inner disc and so the shift to outward migration is likely driven not only by the change in density profile exterior to the planet but also by the factor two increase in surface density interior to the planet during the period of torque reversal. Nevertheless, we find that the simple surface density ratio proposed is a good predictor of the sign of the torque (see \figureautorefname~\ref{Fig:TorqueVSres}).}
{\figureautorefname~\ref{Fig:TorqueVSres} therefore demonstrates that the planet torque is a well-defined function of an {\it output} of the simulation (describing gap shape) but does not itself help to predict what range of {\it input} parameters map onto each outcome. This will be explored in \sectionautorefname~\ref{subsec:Normaliszed torques}.}

{Finally, we investigated the impact of varying the initial and boundary conditions. We found that the initial conditions do not influence the results, whereas the boundary conditions can affect them. In particular, imposing a viscous inflow delivers material more efficiently from the outer disc to the gap region, modifying the disc structure and, consequently, the details of migration (see \appendixautorefname~\ref{app:Varying the initial and boundary conditions} for more details).}

\subsection{Interpretation of inward and outward migration phases}
\label{subsec:Interpretation of inward and outward migration phases}
The fact that the direction of migration is related to the $\Sigma_{1:2}/\Sigma_{1:3}$ ratio implies that the gap structure plays a primary role in determining the direction of migration. Indeed, outward migration takes place when $\Sigma_{1:2}/\Sigma_{1:3}$ drops to values $\lesssim 0.5$, meaning that, in principle, there should be a limiting gap width and/or steepness which triggers outward migration: when both the resonances are located outside the limiting gap width, the density at their location is comparable and the planet migrates inwards; when the 1:2 resonance enters the limiting gap width, the density at the 1:2 resonance drops and the 1:3 resonance dominates, causing planet outward migration. \footnote{We note, however, that at the moment there is no gap-width parameter precise enough to capture whether or not the two resonances are both outside the gap or the 1:2 resonance has shifted inside it.} {All simulations initially undergo inward migration; then, some remain inward, while others transition to outward migration or outward migration followed by stalling.}
\begin{figure*}
    \centering
    \includegraphics[width=0.9\linewidth]{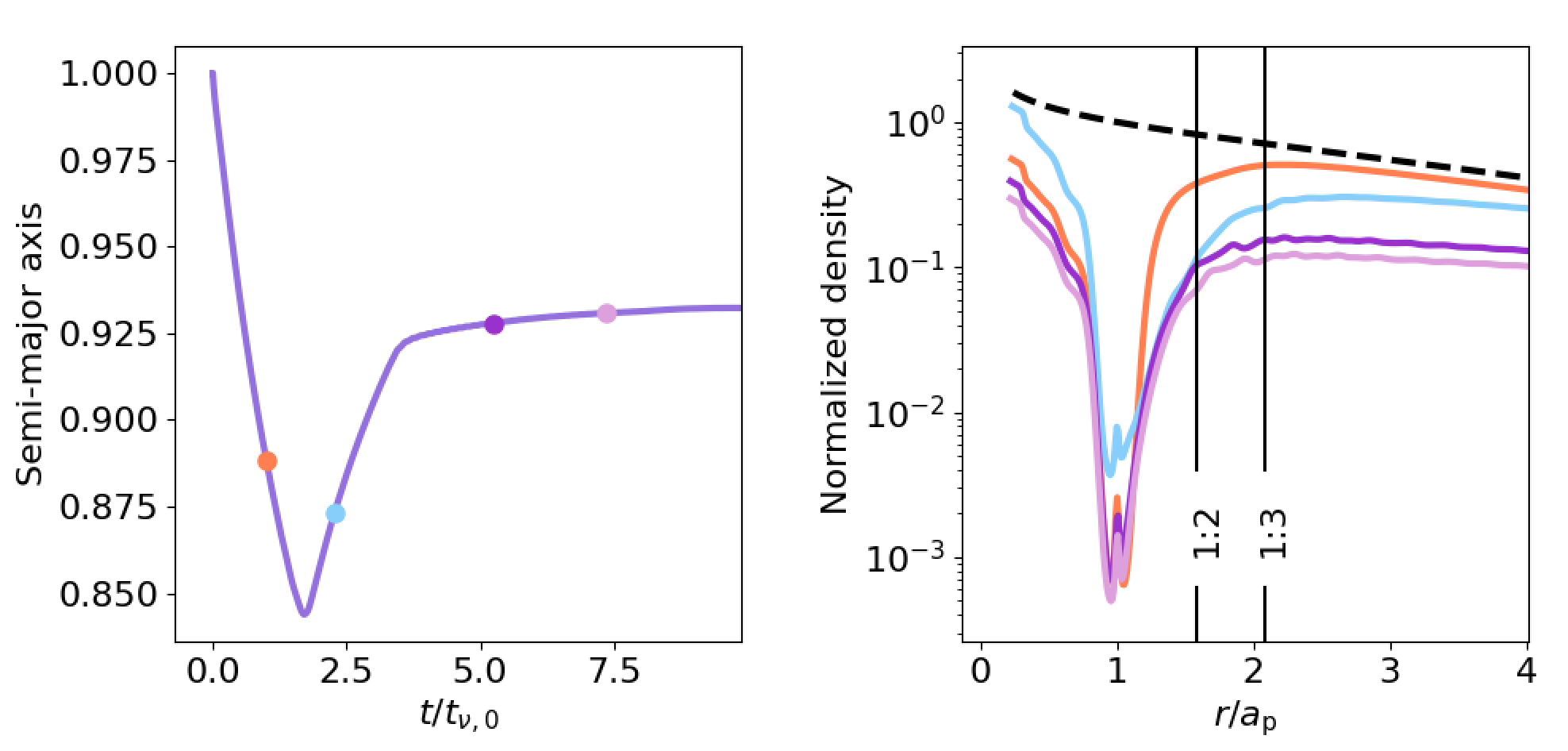}
    \caption{Migration track (left panel) and density profiles (right panel). The colored dots in the left panel show some selected snapshots for inward (orange), outward (light blue) and stalling (purple and pink) phase of migration. The density profiles in the right panel are shown in colors corresponding to the dots in the left panels.}
    \label{Fig:MigPhases}
\end{figure*}

\begin{figure*}
    \centering
    \includegraphics[width=1\linewidth]{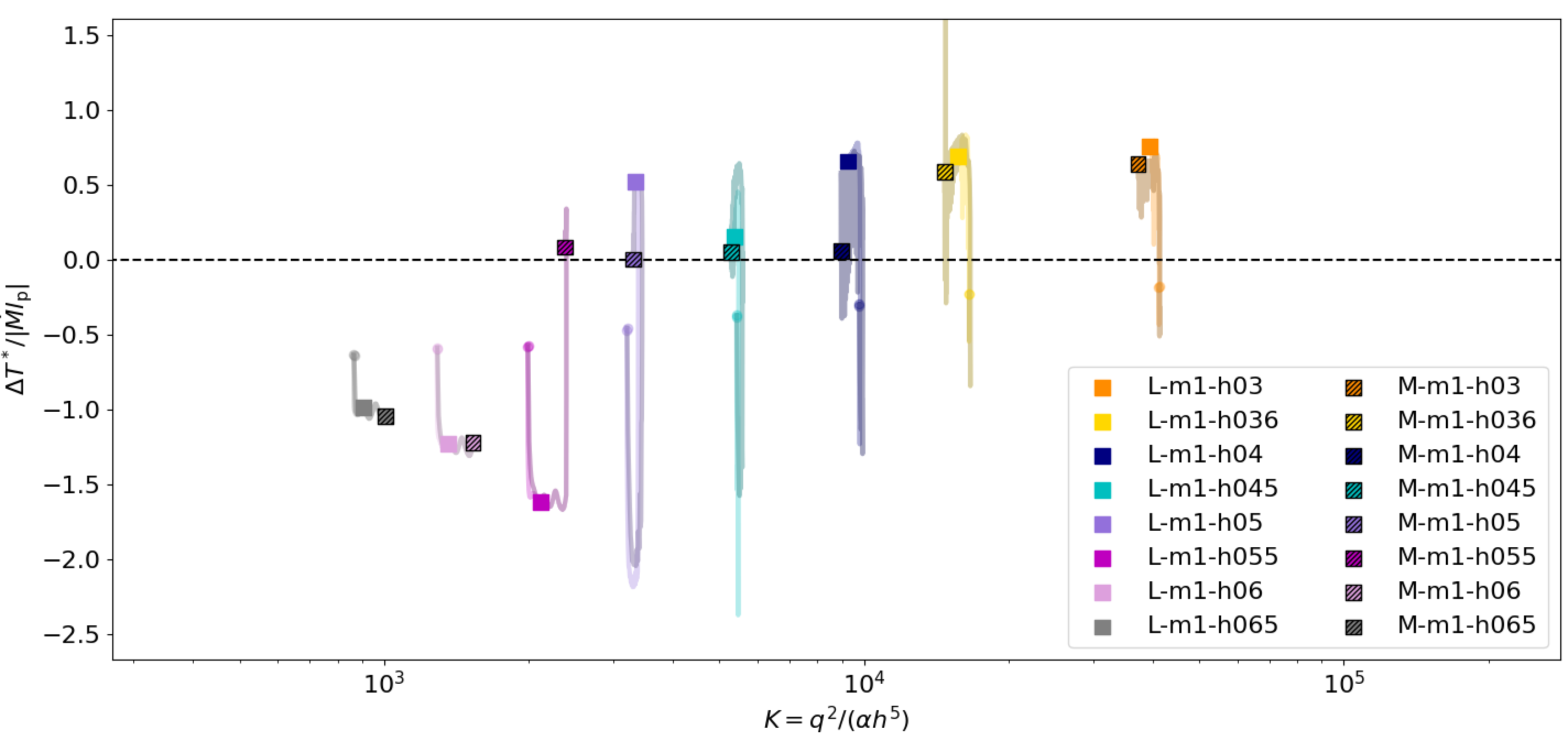}
    \caption{Portion of the normalized torque responsible for the change in the semi-major axis as a function of the gap-shape parameter $K$. Different colors indicate different disc aspect ratios at 5 au, as indicated in the legend. Markers indicate the state of each simulation at $t=10\ t_{\nu,0}$. The lines show the evolution of each simulation. {The high (low) line opacity corresponds to the lower (higher) disc mass regimes.}}
    \label{Fig:TorqueVSk_m1}
\end{figure*}

Given this understanding of migration, we can divide the migration into 3 phases (not all necessarily occurring for all the planets) and interpret them as follows:
\begin{enumerate}
    \item {Inward migration phase}. {The gap is relatively narrow (low $K$)}, and thus both the resonances are outside the gap and the $\Sigma_{1:2}/\Sigma_{1:3}$ ratio is high -- meaning that the planet and disc orbits remain effectively circular. Consequently, the planet migrates inward according to regular planet-dominated Type II migration. The gap profile in this migration phase for our test simulation M-m1-h05 is illustrated by the orange line in the right panel of \figureautorefname~\ref{Fig:MigPhases}.
    \item {Outward migration phase}. The gap is deep (high $K$) and wide enough to reduce the amount of disc material at the 1:2 resonance, as can be seen from the gap profile in the right panel of \figureautorefname~\ref{Fig:MigPhases} (light blue line). This means that the $\Sigma_{1:2}/\Sigma_{1:3}$ ratio becomes low and the outer gap edge becomes eccentric, altering the gap structure. As a consequence, the inner disc torque dominates over the outer disc torque (as now the outer gap is larger due to eccentricity) and the planet migrates outwards.
    \item {Stalling phase}. This phase usually follows an outward migration phase (however small). One possible explanation is that while the planet migrates outwards, it moves closer to the outer gap edge; it is therefore possible that outward migration causes the re-filling of the 1:2 resonance. Meanwhile, the inner gap edge does not lag behind as the planet migrates outward thanks to the material passing through the gap which is increased during outward planet migration and replenishes the inner disc. This restores the equilibrium between the inner and outer contributions to the torque and causes the stalling of the planet. This situation is illustrated by the purple and pink lines in the right panel of \figureautorefname~\ref{Fig:MigPhases}, showing the gap profiles at 2 different times of the stalling phase. {Once a system attains a situation of zero torque, then the planet is stationary thereafter. This means that the disc is not experiencing a time dependent torque from the planet or (to put it another way), the gap shape is not having to continuously readjust to the changing location of the planet and the change in local $h$. This is a `stable point' of the system and subsequent evolution drains down the disc surface density (as can be seen by comparing the purple and pink gap profiles) but does not affect the normalized torque.}
\end{enumerate}

{Although this interpretation can explain these 3 different migration phases, it remains to be understood why the torques always tend to arrange a situation of zero effective torque on the planet.}

\subsection{Identifying an ordering parameter for the sign of planet migration}
\label{subsec:Normaliszed torques}
In \citet{Scardoni+2022} we suggested that the turning point for the torque sign -- previously found in fixed-planet simulations by \citet{Dempsey+2020,Dempsey+2021} -- is also present in live-planet simulations and correlates with the gap depth parameter $K$ (\equationautorefname~\ref{eq:K}). The limited number of simulations used in that work, however, inevitably required further investigation to confirm the presence of a `universal' limiting gap-shape parameter $K_{\rm lim}$ corresponding to the zero torque (and thus the existence of a stalling radius). {We now understand that the direction of migration is set by the relative importance of the 1:3 resonance with respect to the 1:2 resonance: the 1:3 governs eccentricity excitation at the outer gap edge, leading to depletion of material in the outer gap region and outward migration; while the 1:2 promotes circular orbits for the disc material and therefore do not cause any change in the outer gap structure. Consequently, the ratio $\Sigma_{1:2}/\Sigma_{1:3}$ correlates with the direction of migration, with high (low) values favoring inward (outward) migration.} Our interpretation of the different migration phases (\sectionautorefname~\ref{subsec:Interpretation of inward and outward migration phases}) suggests that the gap shape plays an important role in determining $\Sigma_{1:2}/\Sigma_{1:3}$ and thus the migration direction. We therefore explore whether, given the larger parameter space explored in this paper, the gap-depth parameter $K$ is still a good ordering parameter.

\figureautorefname~\ref{Fig:TorqueVSk_m1} shows the evolution of the portion of the torque associated to the change in the semi-major axis, for the subset of simulations with 1 $m_{\rm J}$ planets. The starting point for each simulation is shown by the dot, the final point at 10 $t_{\nu,0}$ is shown by the marker (styles as in the previous plots); the initial and final locations of each simulation are connected via lines that are a slightly  lighter (darker) in color for the light (massive) disc case.

Focusing first on the properties of the simulations at $10\ t_{\nu,0}$, shown by the square markers, we can confirm that $K$ roughly sorts the simulations into those with steady inward migration and steady outward migration. 
However, contrary to what was expected from \citet{Scardoni+2022}, there is no unique location of zero torque - i.e. there is not a single limiting value $K_{\rm lim}$. {Instead there is a range of $K$ ($\sim 10^{3.3}-10^{3.9}$) values over which the system settles to zero torque, making relatively small adjustments to its orbital radius in the process. Outside this range of $K$ the evolution is monotonically inwards or outwards over the duration of the simulation.} The fact that parameter $K$ cannot capture the zero torque location properly is likely because $K$ is controls the gap depth, not the width (which is what causes the reduction of material at the 1:2 resonance). However, since deeper gaps are usually also wider, $K$ is still a relatively good sorting parameter when we consider simulations far from the transition to zero torque. We refer to \sectionautorefname~\ref{subsec:sortingcriterion} for further exploration of the sorting criterion.

We can further notice that, in general, the light and massive versions of each system follow the same torque-K evolution. The light disc cases are less evolved (likely because the lower B value makes planet migration slower), therefore they "lag behind" the massive cases. This suggests that the direction of migration is not influenced by the disc-to-planet mass ratio and therefore this phenomenon is not dependent on a specific history of disc-planet eccentricity exchange. Furthermore, the fact that this happens at different evolutionary times means that the phenomenon is independent of the disc evolution. All of this points in the direction of a local effect, and, therefore a local criterion determining the direction of migration. On the other hand, however, the disc-to-planet mass ratio is still important in determining the distribution of planets after disc dispersal; this is still influenced by the $B$ parameter, as it is determined by the combination of the migration speed and the disc evolution timescale (as discussed in \citealt{Scardoni+2022}). 

{In \sectionautorefname~\ref{sec:Migration of higher mass planets} we explore how well the ordering parameter works as a function of $q$ - i.e. whether the regimes described for $K$ for $1\ m_{\rm J}$ planets work also for higher mass planets.}

\section{Migration of higher mass planets}
\label{sec:Migration of higher mass planets}
After studying the details of migration for Jupiter mass-like planets, we extend our analysis to higher-mass planets in order to explore how the planet mass influences migration. Higher mass planets carve deeper and larger gaps in the disc, thus we expect them to be more prone to outward migration.

{In \figureautorefname~\ref{Fig:TorqueVSres_highmass} we plot the normalized torque as a function of the gap depth parameter $K$ (left panel) and the $\Sigma_{1:2}/\Sigma_{1:3}$ ratio (right panel) for the simulations with higher mass ratio planets. The plot in the left panel confirms that also in the case of higher mass planets, the light and massive version of each simulation have the same evolution (but different evolutionary timescales) in this plane. The right panel, instead, shows some differences with the corresponding plot for Jupiter mass planets (\figureautorefname~\ref{Fig:TorqueVSres}); in particular, 3 simulations (M-m3-h05, L-m13-h10, M-m13-h10) get to the zero torque conditions for values of $\Sigma_{1:2}/\Sigma_{1:3}$ relatively low - which usually corresponds to outward migration.}

{We noticed that the discs hosting $13\ m_{\rm J}$ planets become highly eccentric, with high eccentricity (up to $\sim 0.5$) extending up to $r=5$ and also involving some part of the inner disc.}\footnote{{Note that these behaviors have been verified but the corresponding plots are omitted for conciseness.}} Furthermore, \figureautorefname~\ref{Fig:Ecc_SemiAxis_Evolution} shows that also the planet eccentricity is higher for higher mass planets. Given the different disc structure and planet eccentricity, we believe that these high mass planets are likely entering a different migration regime, as the migration picture emerging from the previous section only applies to systems where the planet remains on a relatively circular orbit and only the outer gap region becomes eccentric. In particular, since the migration regime described in \sectionautorefname~\ref{sec:Migration properties of Jupiter mass-like planets} is only valid for quasi-circular planets migrating in discs that only grow eccentricity close to the outer gap, we need to exclude from our analysis of the inward/outward migration and stalling behavior the following simulations: L-m3-h036, M-m3-h05, L-m13-h036, M-m13-h036, L-m13-h06, M-m13-h06, L-m13-h10, M-m13-h10.\footnote{{We verified that the high eccentricity in these simulations do no cause problematic interaction with the inner boundary by checking that the orbital area of the planet ($a_{\rm p}(1\pm e_{rm p})$) does not get close to the inner boundary over all the simulation run.}}

\begin{figure*}
    \centering
    \includegraphics[width=0.9\linewidth]{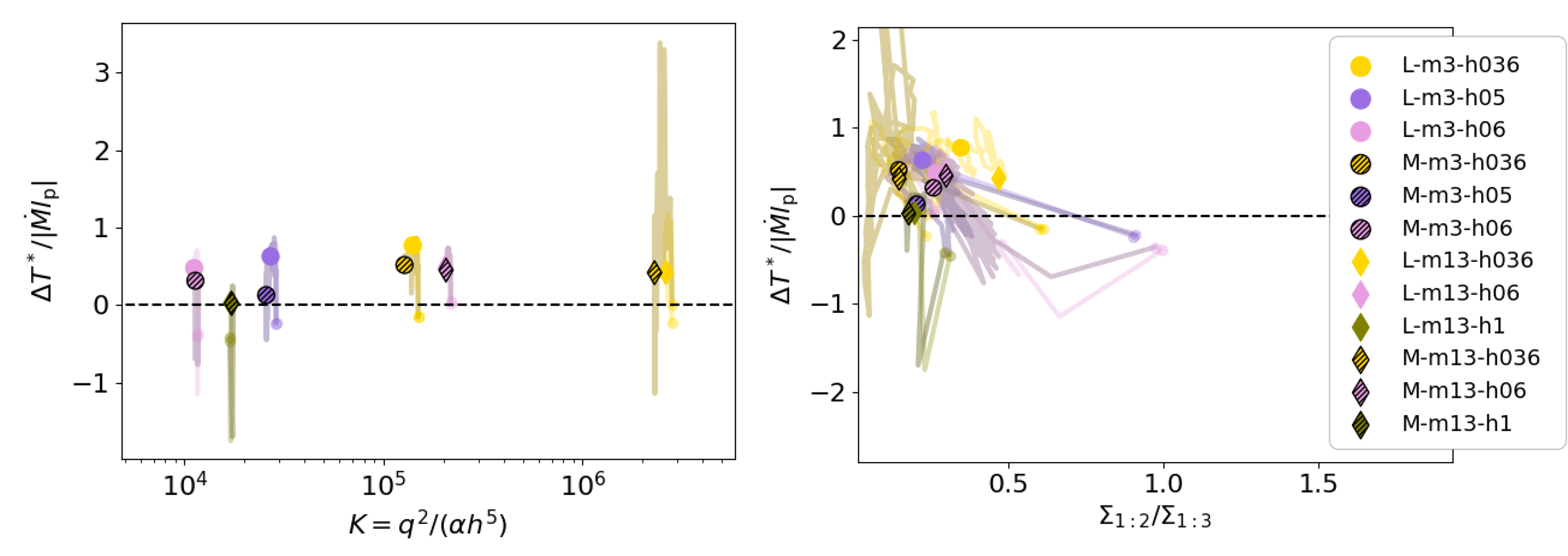}
    \caption{Same as \figureautorefname~\ref{Fig:TorqueVSres} and \figureautorefname~\ref{Fig:TorqueVSk_m1} but for higher mass planets.}
    \label{Fig:TorqueVSres_highmass}
\end{figure*}

\begin{figure*}
    \centering
    \includegraphics[width=1\linewidth]{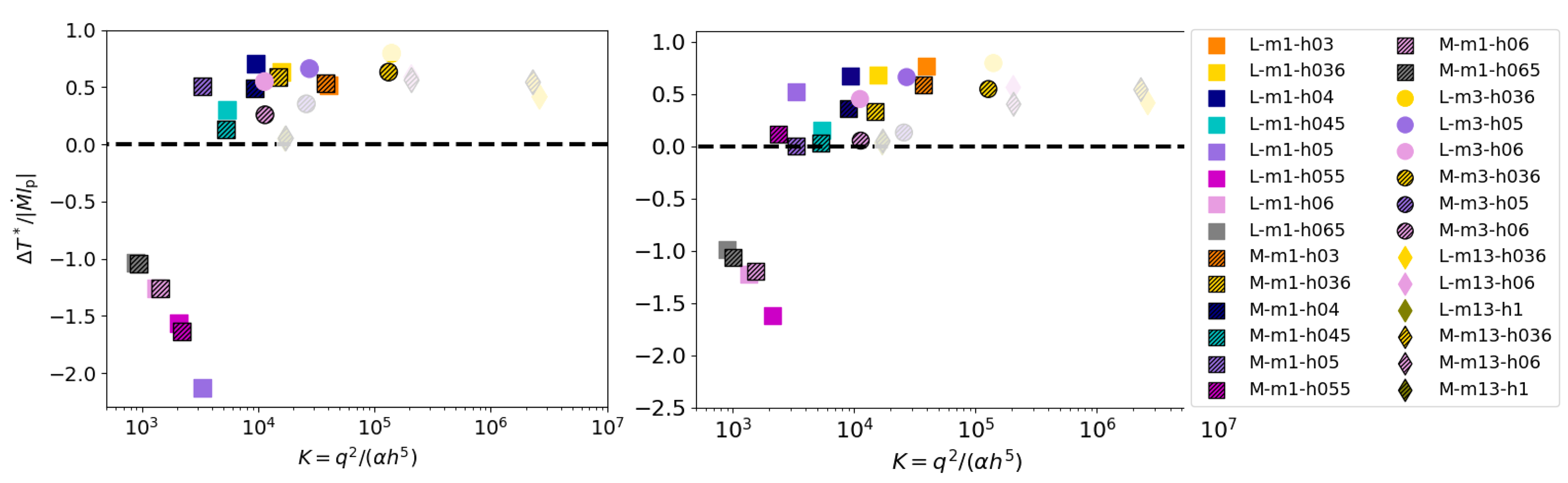}
    \caption{Portion of the normalized torque responsible for the change in the semi-major axis as a function of the gap-shape parameter $K$. In the left panel, the plot shows the distribution of the simulations at $t=3\ t_{\nu,0}$, while the right panel shows the distribution at $t=10\ t_{\nu,0}$. 
    The marker shape indicates the planet mass: 1 $m_{\rm Jup}$ (squares), 3 $m_{\rm Jup}$ (circles), 13 $m_{\rm Jup}$ (diamonds). The colors indicate the disc aspect ratio at 5 au. Dashed markers indicate the massive disc simulations. The translucent markers indicate the systems entering a different migration regime with respect to that studied in this paper.}
    \label{Fig:TorqueVSk}
\end{figure*}

In \figureautorefname~\ref{Fig:TorqueVSk} we show the normalized torque associated with the change in the semi-major axis (at 3 $t_{\nu,0}$ and 10 $t_{\nu,0}$ in the left and right panel, respectively) as a function of the gap shape parameter for the entire set of simulations. The systems entering a different migration regime are shown with translucent markers. This plot confirms that planets in systems characterized by high (low) $K$ values migrate outward (inward), even when we consider higher mass planets, provided that we only consider the systems migrating according to the migration regime described in \sectionautorefname~\ref{sec:Migration properties of Jupiter mass-like planets} {(i.e. circular planet orbits and moderate disc eccentricity)}.
{From \figureautorefname~\ref{Fig:TorqueVSk_m1} and the left panel of \figureautorefname~\ref{Fig:TorqueVSres_highmass} we understand that each system is undergoing its own little loop in $\Delta T^* /|\dot{M}l_{\rm p}|$ versus $K$ space on its own timescale; therefore the big scatter in torques at a given $K$ value in the range $\sim2000-10000$ in \figureautorefname~\ref{Fig:TorqueVSk} (only considering simulations in the circular planet, eccentric outer gap migration regime) reflects different timescales in performing the same loop. }

\section{Discussion}
\label{sec:Discussion}
\subsection{Disc and planet parameters for outward migration}
\label{subsec:Disc and planet parameters for outward migration}
Although not being the ideal parameter to sort the inward and outward migration around the point where the torques switch their sign, $K$ is still a good sorting parameter for inward and outward migration far from $\Delta T/|\dot{M}l_{\rm p}|=0$. In \figureautorefname~\ref{Fig:hVSmp_K} we therefore explore the $K$ map as a function of the local disc aspect ratio $h$ ($y$ axis) and planet mass $m_{\rm p}$ ($x$ axis) for $\alpha=10^{-3}$ (as that considered in our set of simulations). The brown-yellow area indicates the combination of $h,\ \alpha,\ m_{\rm p}$ that can produce inward migration; the teal area shows the combination of parameters that is expected to correspond to outward migration. The area between $\log(K) = 3.3$ and $\log(K) = 4$ (indicated by gray lines) is the region where there is a significant scatter in the values of the normalized torque. As expected, higher mass planets are more prone to outward migration with respect to the lower mass planets.\footnote{In principle, lower values of the $\alpha$ viscosity parameter tend to favor outward migration, while higher values favor inward migration. However, because we did not test different $\alpha$ values and the migration direction depends strongly on the gap shape, this expectation requires confirmation through further simulations.} Nonetheless, these plots show that reasonable disc parameters can produce outward migration for 1 $m_{\rm Jup}$ - this means that outward migration is not hard to obtain in this mass regime.
\begin{figure}
    \centering
    \includegraphics[width=1\linewidth]{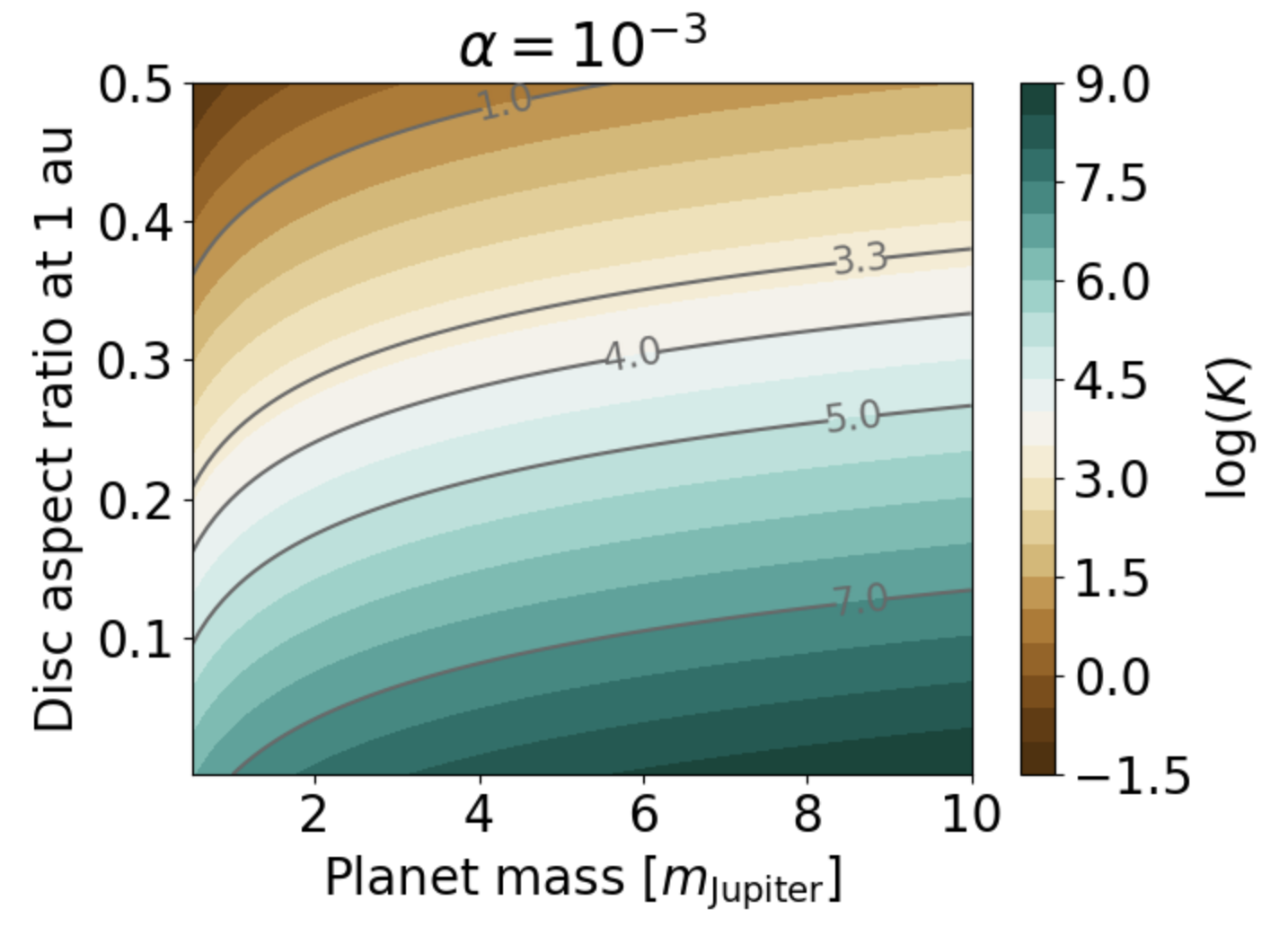}
    \caption{Gap-shape parameter $K$ map as a function of the local disc aspect ratio (y axis) and planet mass (x axis). Some reference values for $K$ are highlighted via the gray contours (the "stalling region" is within the interval $\log(K)=[3.3,4.1]$). The brown-yellow area indicates the combination of $h,\ \alpha,\ m_{\rm p}$ that is expected to generate inward migration; the teal area shows the combination of parameters that is expected to correspond to outward migration.}
    \label{Fig:hVSmp_K}
\end{figure}

\subsection{Stalling radius and comparison with the toy model}
\label{subsec:Stalling radius}
\begin{figure*}
    \centering
    \includegraphics[width=1\linewidth]{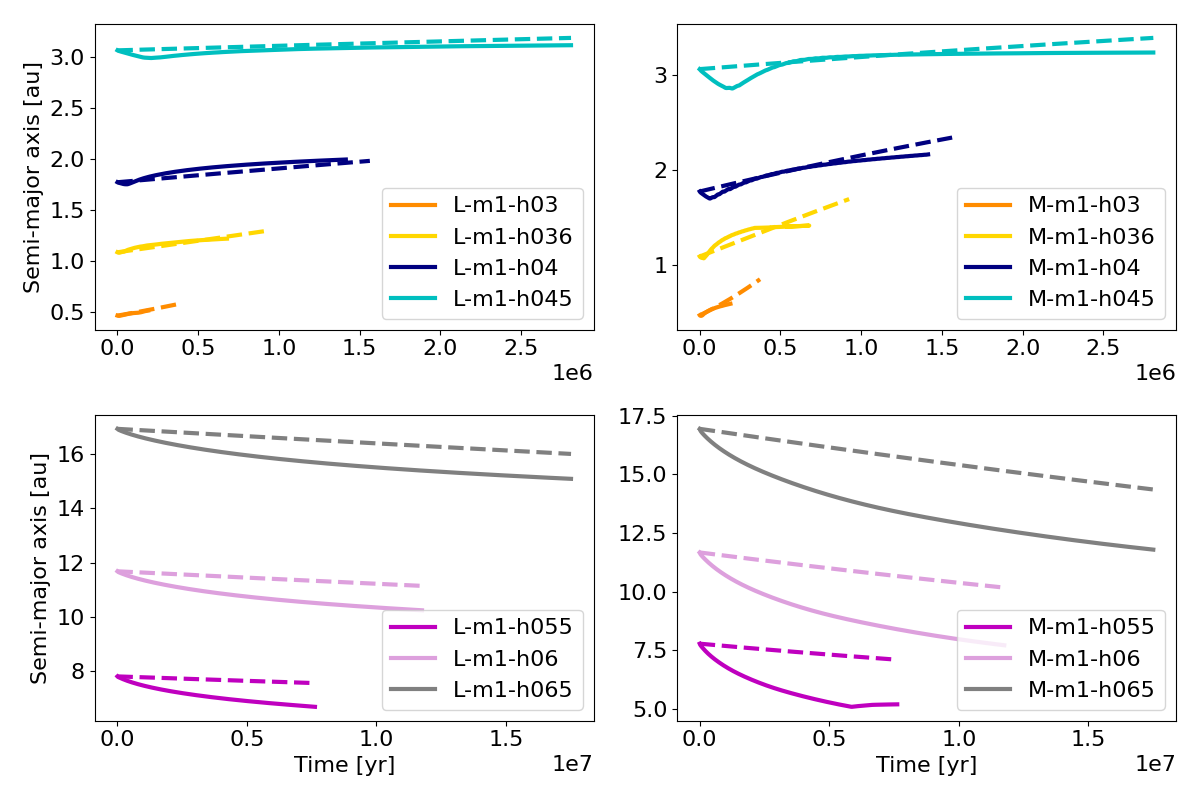}
    \caption{Evolution of Jupiter-mass planets in physical units for light (left) and massive (right) discs. Solid lines show simulation tracks, while dashed lines indicate predictions from the toy model of \citet{Scardoni+2022} ($W=1$, $K_{\rm lim}=3500$).}
    \label{Fig:evolution_physicalunits}
\end{figure*}
In our approximate model, based on the idea that the normalised torque correlates with $K$, the stalling radius is defined as (see also \citealt{Scardoni+2022})
\begin{equation}
    r_{\rm lim}=\left(\frac{q^2}{K_{\rm lim}\alpha_0 h_0^5}\right)^{1/(a+5f)}.
    \label{eq:rlim}
\end{equation}
{However, given the uncertainties in $K_{\rm lim}$, predicting the exact stalling radius remains challenging. In \figureautorefname~\ref{Fig:evolution_physicalunits}, we show the evolution of a subset of simulations with Jupiter-mass planets in physical units. To convert from code units to physical units, we assumed a disc with $h=0.05$ at 5 au (corresponding to simulation M-m1-h05), which translates to $h \sim 0.035$ at 1 au. The simulation tracks are shown as solid lines for the light and massive disc cases in the left and right panels, respectively. From this plot we note that, in the selected disc model, the typical stalling radius stalls between 3–5 au (see the cyan and magenta lines). If we consider our range of $h=0.03-0.065$ at 5 au (corresponding to $\sim 0.02-0.05$ at 1 au) the stalling radius ranges from 0.5 au to 50 au; this underlines that the stalling radius is quite sensitive to the aspect ratio (indeed, the gap properties are sensitive to $h$).}

{Furthermore, we indicate with dashed lines the predicted evolution according to the toy model of \citet{Scardoni+2022} (with $W=1$ and $K_{\rm lim}=3500$), which provides an analytic estimate of the migration track based solely on the gap-shape parameter. While the toy model does not capture the full complexity of gap eccentricity and long-term disc evolution, it nonetheless reproduces the qualitative behavior observed in our simulations, including the occurrence of outward migration and stalling at radii of a few au. This highlights the need for a precise gap profile model in order to be able to predict the migration precisely.} {We notice that the toy model captures the migration tracks of outward-migrating planets quite well, whereas it tends to predict slower migration than observed in the simulations for inward-migrating planets. This discrepancy is likely due to a transient phase in which inward-migrating planets move faster than the velocity prescribed by Type II migration (see \citealt{Scardoni+2020}). At later times, the migration rate tends to slow to values more consistent with the prediction (this can be seen evidently from the gray lines, which evolved for longer than the pink and magenta cases). Nonetheless, the initial fast transient is significant enough to make the prescription for inward-migrating planets inaccurate.}

\subsection{Alternative criteria for sorting the simulations according to their direction of migration}
\label{subsec:sortingcriterion}
So far, we have discussed the $\Delta T/|\dot{M}l_{\rm p}|-K$ relation, noticing that the gap-depth parameter is a relatively good parameter to describe and predict the direction of migration. As noticed in \sectionautorefname~\ref{subsec:Interpretation of inward and outward migration phases}, however, the direction of migration is actually driven by the relative importance of the 1:2 and 1:3 outer resonances; therefore, the ideal parameter to predict the direction of migration would be a gap width parameter able to determine whether the 1:2 resonance is located inside or outside the gap. However, obtaining such a parameter would require precise knowledge of the gap shape. The closest parameter available at the moment is the gap-width parameter $K'=q^2/(\alpha h^3)$ derived by \citet{Kanagawa+2015}. We tested the ability of $K'$ to predict the direction of migration in \figureautorefname~\ref{Fig:TorqueVSKprime}; although migration behavior can be efficiently predicted for high and low values of $K'$, also $K'$ cannot predict precisely the zero torque (or the direction of migration turning point). This means that the $K'$ parameter is probably not precise enough to describe the gap-width for this particular problem.

For the moment, we therefore believe that the best available parameter to predict the direction of migration remains $K$, given the lower relative uncertainty at the turning point for the direction of migration. However, we caution that it is not able to capture the details of migration close to the turning point. We believe that a precise prediction of the direction of migration necessarily requires a deep investigation of the gap shape.

\begin{figure}
    \centering
    \includegraphics[width=1\linewidth]{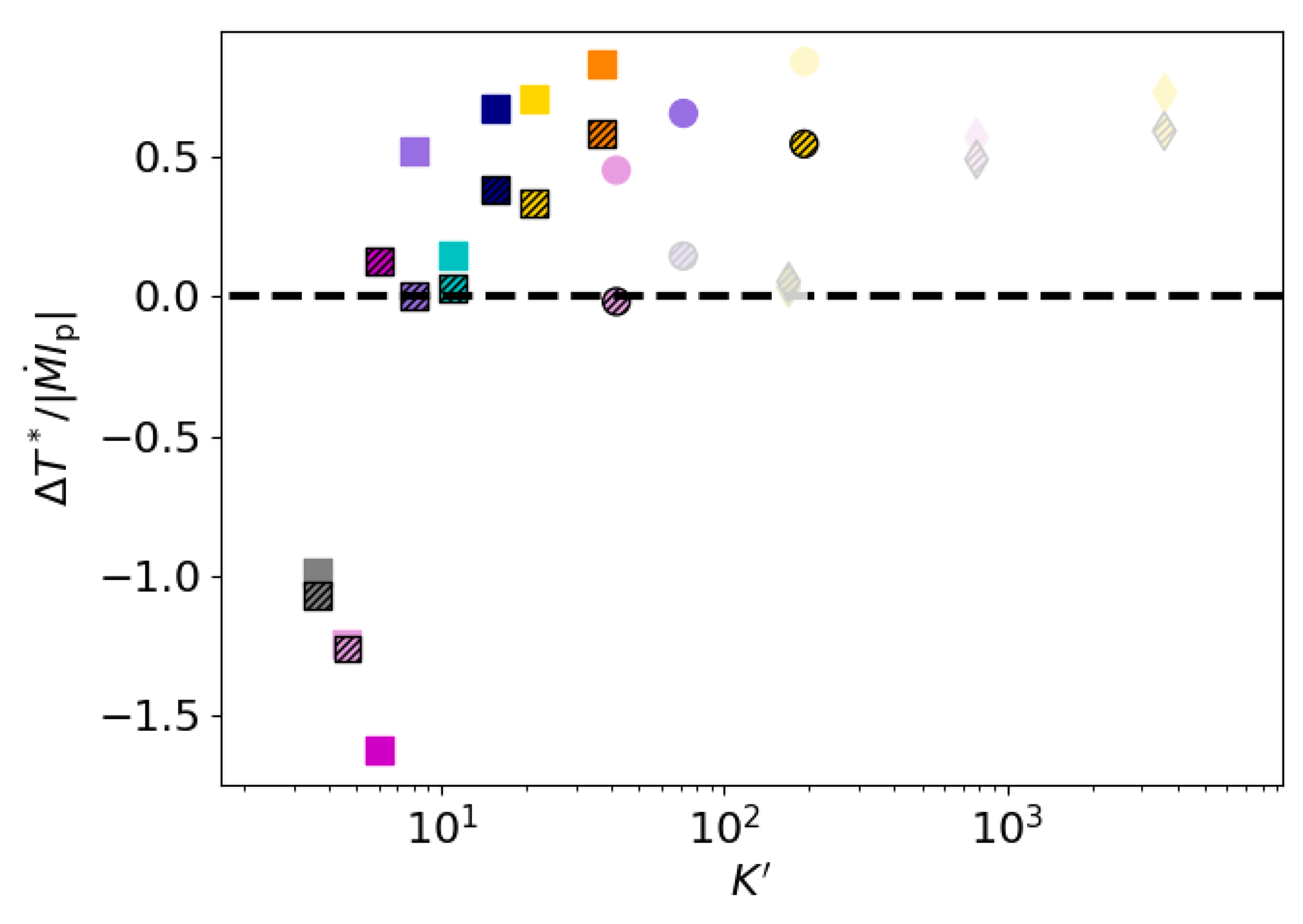}
    \caption{Normalized torque as a function of the gap-width parameter $K'=q^2/\alpha h^3$.}
    \label{Fig:TorqueVSKprime}
\end{figure}

\section{Conclusions}
In this work, we investigated the migration of massive planets in the planet-dominated regime of Type II migration, focusing on the transition between inward and outward migration. Building upon the torque framework introduced in \citet{Scardoni+2022}, we have run a suite of simulations that explore a wide range of planet masses and aspect ratios, while keeping the $\alpha$ viscosity parameter fixed.

Our main findings can be summarized as follows:
\begin{enumerate}
    \item A critical insight emerging from our analysis is the key role of the 1:2 and 1:3 outer Lindblad resonances in determining the migration direction (\sectionautorefname~\ref{subsec:The role of the disc structure and eccentricity in outward planet migration} and \sectionautorefname~\ref{subsec:Interpretation of inward and outward migration phases}). The relative positioning of these resonances with respect to the outer gap edge strongly influences their contribution to the torque acting on the planet, and therefore the overall torque balance (and hence the migration outcome). Specifically, at the beginning of the simulations both resonances lie outside the gap, thus the 1:2 resonance dominates, the disc remains circular, and the planet migrates inwards. While the system evolves, in some cases, the 1:2 resonance shifts inside the gap region. Consequently, its contribution weakens, allowing the 1:3 resonance, that remains in the high density region outside the gap, to dominate. When the 1:3 resonance dominates, it causes an increase in the eccentricity of the disc outer gap region and favors the gas passage through the gap. These effects significantly alter the gap structure, causing the density to increase in the inner disc and to decrease in the outer gap region, leading to outward migration. Finally, during the outward migration phase, the outer gap eccentricity tends to decrease and the gap shape tends to assess an equilibrium, causing the planet to stall.
    \item In \sectionautorefname~\ref{subsec:Normaliszed torques} we confirmed that the gap-depth parameter $K=q^2/(\alpha h^5)$ is a reasonably effective predictor of the migration direction, with outward migration favored for higher $K$ (i.e. higher planet masses, lower viscosity, thinner discs) - as already suggested by \citet{Scardoni+2022}. However, we found that while $K$ performs well far from the torque transition region $2500 < K < 10000$, it becomes less reliable near the turning point, where the migration direction switches sign. {The stalling radius (i.e. the orbital location corresponding to the zero torque condition), which depends on a critical value of $K$, also remains uncertain because the torque reversal and stalling can occur over a broad range of $K$ values.} We tested the gap-width parameter $K'$ as an alternative, but found it less effective in predicting the turning point of migration (\sectionautorefname~\ref{subsec:sortingcriterion}). This implies that the precise location of the migration turning point remains uncertain, as simple gap parameters cannot capture the migration behavior close to the zero torque condition. We suggested that the best parameter to predict the direction of migration would be a precise gap-width parameter, but this is not available at the moment.
    \item Our findings suggest that outward migration is not only obtained for extreme disc conditions but can occur for Jupiter-mass planets in reasonable disc environments (\sectionautorefname~\ref{subsec:Disc and planet parameters for outward migration} and \sectionautorefname~\ref{subsec:Stalling radius}).
    \item In \sectionautorefname~\ref{sec:Migration of higher mass planets} we extended our analysis to higher mass planets, finding that this new migration regime regulated by the 1:2 and 1:3 outer resonances is only valid for systems with {near circular orbit planets} and discs with relatively high eccentricity in outer gap region; in cases where the planet becomes significantly eccentric or the disc eccentricity involves large areas of the disc (including the inner disc or outer disc far away from the gap), the system enters a different regime and cannot be described by our analysis.
\end{enumerate}

While the results presented in this paper provide valuable insights into the migration behavior of massive planets, several limitations and unresolved issues remain. Below, we outline key caveats in our current approach, as well as open questions:
\begin{itemize}
    \item The $\Sigma_{1:2}/\Sigma_{1:3}$ ratio remains descriptive rather than predictive of the direction of planet migration, and the gap-depth parameter $K$ does not reliably capture migration behavior near the zero-torque condition. A precise prediction of the migration turning point requires a more accurate characterization of the gap structure.
    \item The uncertainties in torque prescriptions near the transition regime must be carefully considered when applying these results to population synthesis models or interpreting observations.
    \item It remains unclear why planets in this migration regime systematically evolve toward the equilibrium condition (i.e., zero net torque).
\end{itemize}

\begin{acknowledgements}
We thank the referee for a careful reading of the manuscript and helpful comments. We thank Alessandro Morbidelli and Elena Lega for valuable discussions that significantly contributed to this study.
C.E.S and G.P.R. acknowledge support from the European Union (ERC Starting Grant DiscEvol, project No. 101039651) and from Fondazione Cariplo, grant No. 2022-2017. Views and opinion expressed are, however, those of the author(s) only and do not necessarily reflect those of the European Union or European Research Council. Neither the European Union nor the granting authority can be held responsible for them.
CJC has been supported by the UK Science and Technology research Council (STFC) via the consolidated grant ST/W000997/1.
ER acknowledges financial support from the European Union's Horizon Europe research and innovation programme under the Marie Sk\l{}odowska-Curie grant agreement No. 101102964 (ORBIT-D).
RAB thanks the Royal Society for their support through a University Research Fellowship.
This work was in part performed using the Cambridge Service for Data Driven Discovery (CSD3), part of which is operated by the University of Cambridge Research Computing on behalf of the STFC DiRAC HPC Facility (www.dirac.ac.uk). The DiRAC component of CSD3 was funded by BEIS capital funding via STFC capital grants ST/P002307/1 and ST/R002452/1, and STFC operations grant ST/R00689X/1. This work was in part performed at CINECA, with computational resources on the LEONARDO supercomputer, owned by the EuroHPC Joint Undertaking and hosted by CINECA (Italy).
\end{acknowledgements}

\bibliographystyle{aa} 
\bibliography{references} 

\begin{appendix} 
\onecolumn

\section{Evolution of the 1:3 vs 1:3 resonances}
\label{app:Evolution of the 2:1 vs 3:1 resonances}
In \figureautorefname~\ref{Fig:Resonances} the colored lines show the $\Sigma_{1:2}/\Sigma_{1:3}$ ratio as a function of the evolutionary time $t/t_{\nu,0}$ for all simulations with 1 Jupiter mass planets. The light gray line in each plot shows the rate of change of the planet semi-major axis $\dot{a}_{\rm p}/|\max(\dot{a}_{\rm p})|$.

We notice a strong correlation between the value of $\Sigma_{1:2}/\Sigma_{1:3}$ ratio and $\dot{a}_{\rm p}/|\max(\dot{a}_{\rm p})|$: inward (outward) migration phases are associated with high (low) values for $\Sigma_{1:2}/\Sigma_{1:3}$.\footnote{Note that $\Sigma_{1:2}/\Sigma_{1:3}$ is initially very high for all the simulations because of the initial conditions, where the planet is not present, but it is gradually inserted over the first 50 orbits of the simulation.} This suggests that the 1:3 resonance, and therefore $\Sigma_{1:2}/\Sigma_{1:3}$, is key in the outward migration process. Similar results are found for all the other simulations in this migration regime (L-m3-h036, M-m3-h036, L-m3-h05, M-m3-h05, M-m3-h06).
\begin{figure*}[h!]
    \centering
    \includegraphics[width=1\linewidth]{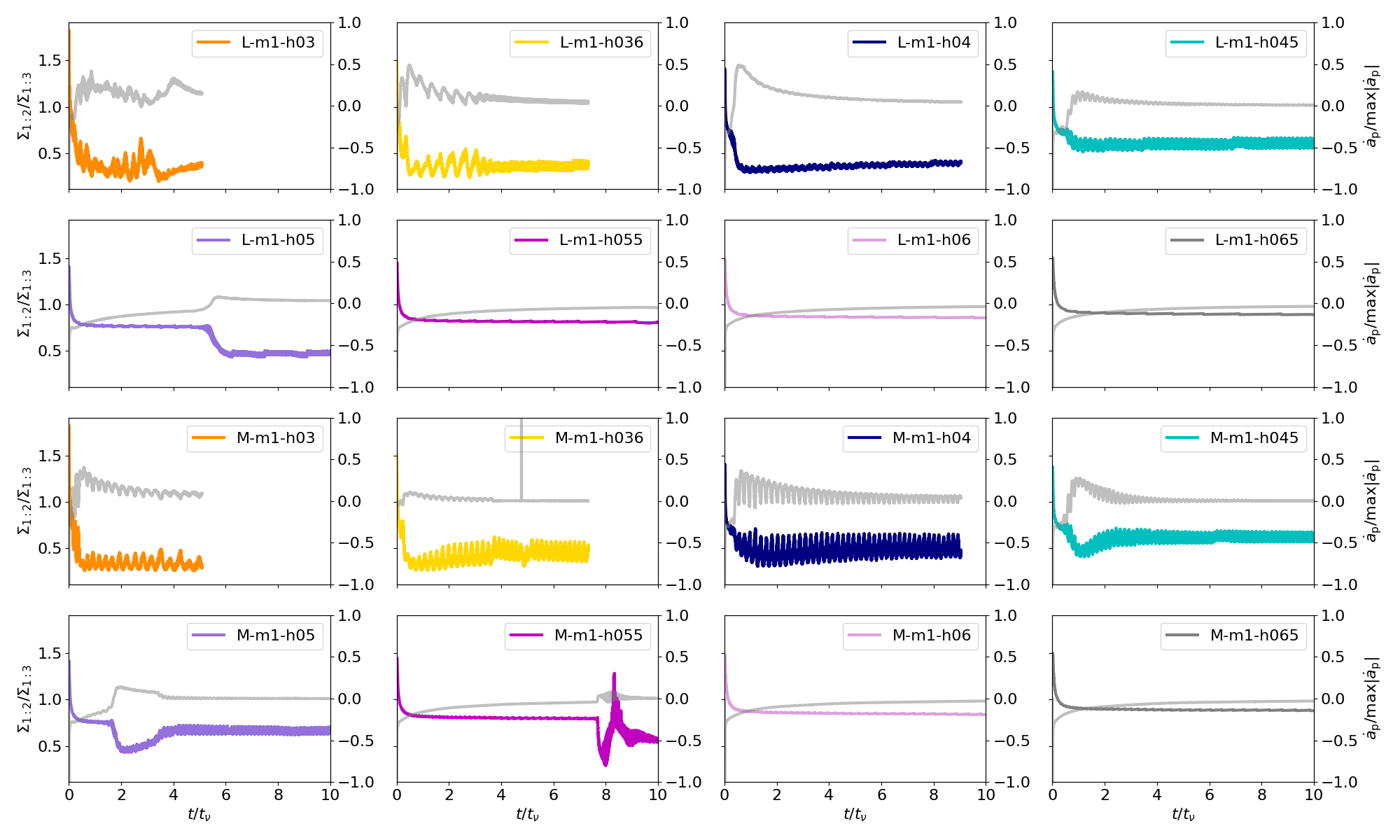}
    \caption{Ratio between the disc density at the 1:2 resonance and at the 1:3 resonance for all simulations with 1 $m_{\rm Jup}$ planets (colored lines), as indicated in each panel legend. The gray lines in each plot show the rate of change of the planet semi-major axis $\dot{a}_{\rm p}/|\max(\dot{a}_{\rm p})|$.}
    \label{Fig:Resonances}
\end{figure*}

\newpage
\section{Varying the initial and boundary conditions}
\label{app:Varying the initial and boundary conditions}
In this appendix we check whether the initial or boundary conditions influence the outcome of migration and the stalling radius. For this purpose, we take as reference simulation M-m1-h05, and perform simulations  M-m1-h05\_VSS and M-m1-h05\_testIC with different boundary conditions and initial conditions, respectively.

In simulation M-m1-h05\_VSS we test the boundary conditions by changing the outer boundary. In our standard configuration, the disc extends up to $r=15$, with an exponential taper starting at $r\sim 5$; for the boundary condition test, we take a viscous steady state disc with a steady state outer boundary condition. The results from this test are shown in by the red line both in \figureautorefname~\ref{Fig:ci_tau5a_taper_visc_jup_H5_testIC_a} and in \figureautorefname~\ref{Fig:ci_tau5a_taper_visc_jup_H5_testIC_b}. 

To test the effect of initial conditions we change the planet initial orbital radius from at $r=1$ (in the standard configuration) to $r=0.84$ (in the initial condition test). The results are shown by the grey line in both \figureautorefname~\ref{Fig:ci_tau5a_taper_visc_jup_H5_testIC_a} and \figureautorefname~\ref{Fig:ci_tau5a_taper_visc_jup_H5_testIC_b}. For reference, the equivalent standard simulation is shown in both the figures by the purple line.

\figureautorefname~\ref{Fig:ci_tau5a_taper_visc_jup_H5_testIC_a} shows the semi-major axis as a function of radius. From this figure we notice that the stalling radius is somewhat sensitive to the outer boundary condition, and it is smaller than in the standard case - this is expected because the viscous outer boundary condition suppresses outward radial flow (which is instead present in the closed outer boundary condition) and therefore increases the dominance of the outer disc to the net torque.\footnote{{For closed boundary conditions, outward radial flow occurs because the system’s material can freely adjust according to viscosity. By contrast, with a forced viscous outer boundary, the inward viscous flow is imposed, so no radial outflow is expected.}} Instead, the stalling radius does not depend on the initial condition, confirming that $r_{\rm lim}$ is an intrinsic location that only depends on the disc ($h$, $\alpha$, $B$) and planet ($q$) parameters and it is not influenced by the specific history of migration.
\begin{figure}
    \centering
    \includegraphics[width=0.4\linewidth]{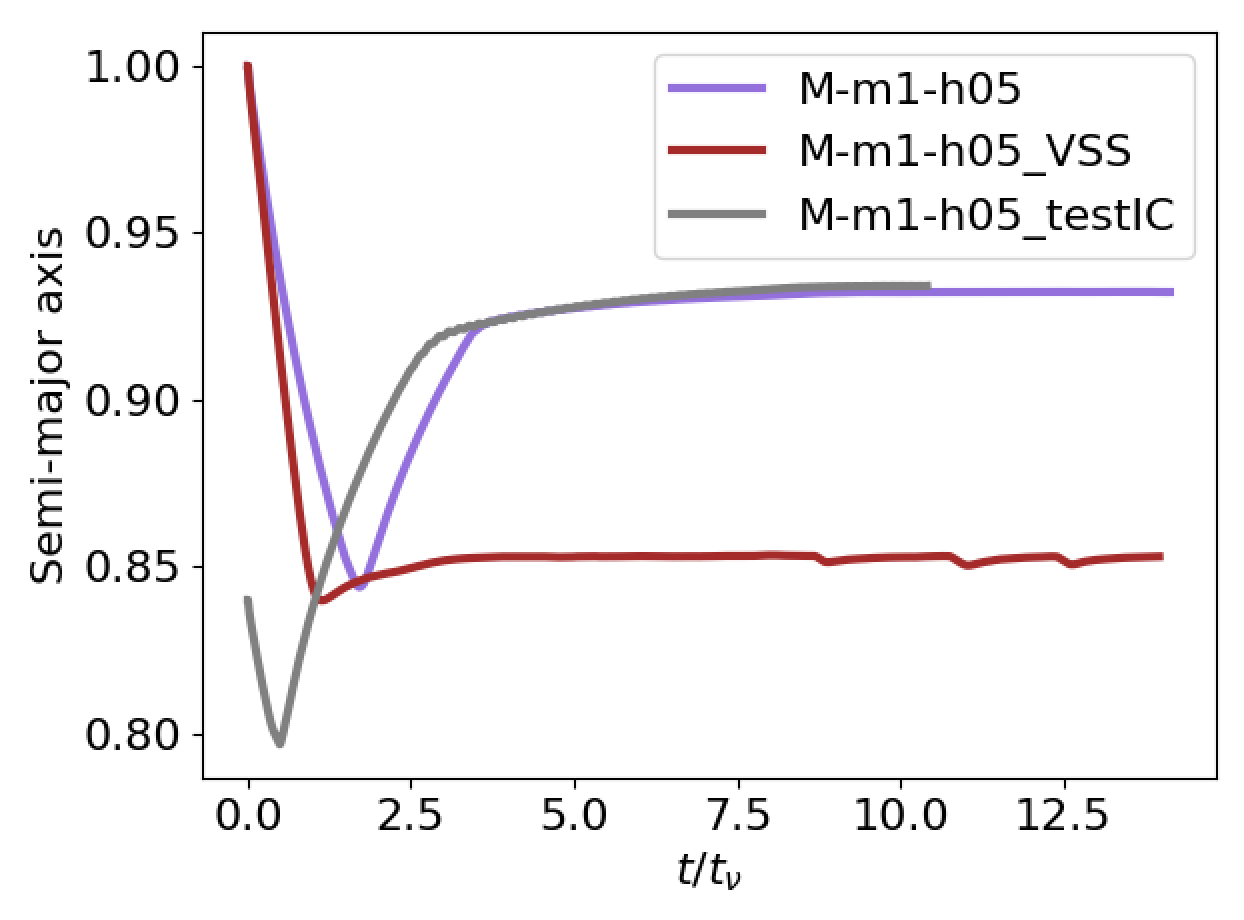}
    \caption{Migration tracks for the system M-m1-h05 in the standard configuration (purple line); with different boundary conditions (red line); with different initial conditions (gray line).}
    \label{Fig:ci_tau5a_taper_visc_jup_H5_testIC_a}
\end{figure}

\figureautorefname~\ref{Fig:ci_tau5a_taper_visc_jup_H5_testIC_b} shows the density ratio at the 1:2 and 1:3 outer resonances, showing that the migration behavior illustrated in the paper is valid also for different initial and boundary conditions.
\begin{figure*}
    \centering
    \includegraphics[width=0.8\linewidth]{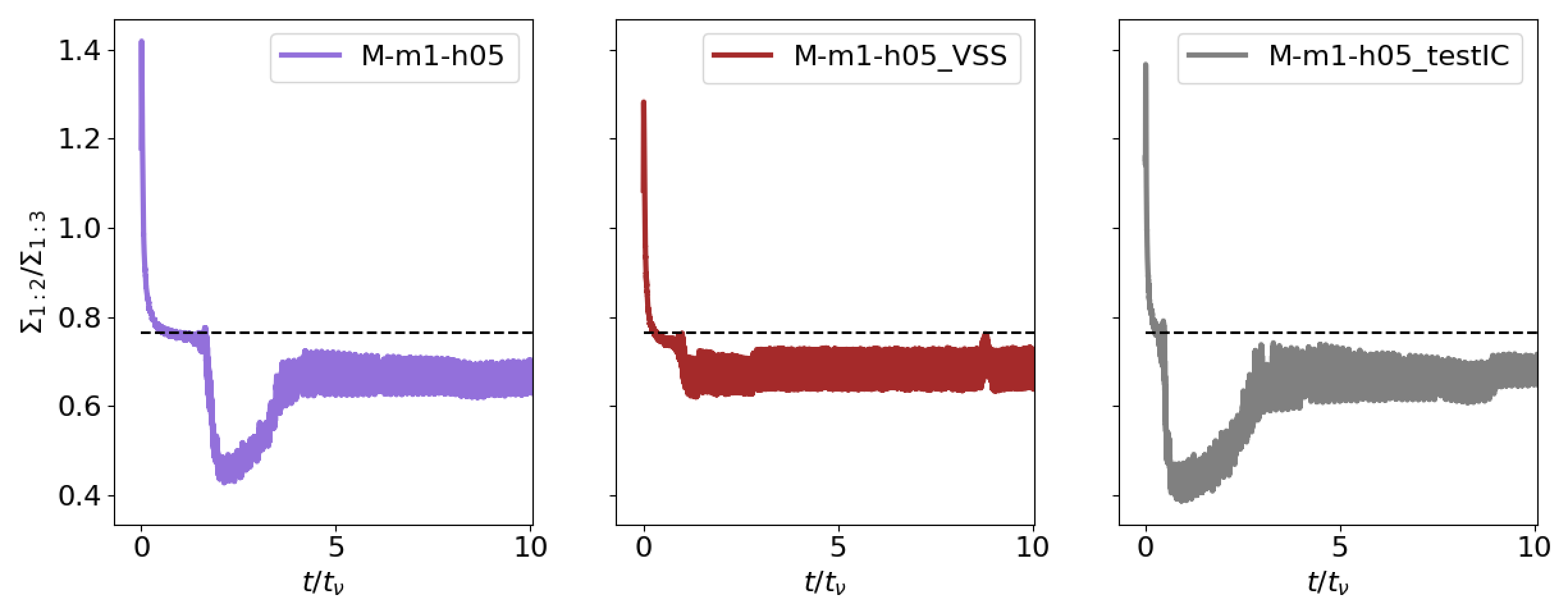}
    \caption{Density at resonances for the system M-m1-h05 in the standard configuration (purple line); with different boundary conditions (red line); with different initial conditions (gray line).}
    \label{Fig:ci_tau5a_taper_visc_jup_H5_testIC_b}
\end{figure*}

\end{appendix}

\end{document}